 \documentclass[pre,showpacs]{revtex4}
\usepackage{graphicx}
\usepackage{bm}

\begin{document}
\topskip 20mm
 \title{Eigenfunction Statistics of Complex Systems: A Common Mathematical Formulation}
 \author{Pragya Shukla}
 \affiliation{Department of Physics,
 Indian Institute of Technology, Kharagpur, India.}
 \date{\today}
 \widetext

 \begin{abstract}
         We derive a common mathematical formulation 
 for the eigenfunction statistics of Hermitian operators, represented by a 
 multi-parametric probability density. The system-information in the 
 formulation enters through two parameters only, namely, system size and 
 the complexity parameter, a function of all system parameter including size. 
 The behavior is contrary to the eigenvalue statistics which is sensitive to 
 complexity parameter only and shows a single parametric scaling.
   
 The existence of a mathematical formulation, of both eigenfunctions and eigenvalues, 
 common to a wide range of complex systems indicates the possibility of a similar 
 formulation for many physical properties. This also suggests the possibility to 
 classify them in various universality classes defined by complexity parameter. 

 \end{abstract}

 \pacs{  PACS numbers: 05.45+b, 03.65 sq, 05.40+j}
 

 \maketitle

 \section {Introduction}
 
The eigenfunction correlations of various generators of dynamics  contain 
 a wealth of information about the system e.g. localized or delocalized 
 nature of the dynamics, decay rate etc. 
 Recently the correlations were shown to be relevant for 
 description of fluctuations of physical properties e.g. conductance in 
 mesoscopic systems, peak-height statistics in coulomb blockade regime 
 of quantum dots \cite{alh,sd}.
 The correlations may vary from level to level or fluctuate in different 
 realizations of a complex system. The strong fluctuations of eigenfunctions are 
 already known to be the hallmark of many critical point studies e.g. 
 metal-insulator transition in disorder systems \cite{mir}, spin glass \cite{ay}, 
  and stock market fluctuations \cite{pl} etc. 
 Recent studies have revealed existence of the fluctuations in 
 a wider range of complex systems e.g. in the area of quantum information, 
 nanotechnology\cite{sd} and complex networks etc. \cite{da}. As a consequence, 
 a detailed information about the eigenfunction statistics of complex systems is 
 very important and desirable.
 
During recent years, many attempts have been made to statistically formulate the 
eigenfunction  correlations of complex systems; see, for example,   
\cite{alh, mir, me, brody, fren, gu, fyd, fe, pg, ap, fal, mmms, mj, berry, psz}. 
One of the main tools used in this context is the random matrix approach which 
can briefly be described as follows (see \cite{me} for details). The  
presence of complicated interactions (among its various sub-units) in the system 
under investigation  often makes it impossible to exactly determine the 
relevant operator e.g. in a matrix representation. The elements (some or all) 
of the operator in the representation can then be best described by a 
probability distribution. 
This permits one to replace the operator by an ensemble of the operators which 
is supposed to describe the generic properties and is referred as 
 random matrix model of the operator. (In this paper, we focus on   
systems where complicated interactions, of any origin, lead to a 
partial or full randomization, of the operator, thus, allowing one to use a random 
matrix approach).    

 The choice of an appropriate random matrix model for a system is  
sensitive to its physical conditions (i.e nature and degree of interactions in 
various sub-units, symmetry and topological conditions, dimensionality etc.). 
This is because the  distribution parameters of each matrix element depend on the 
interactions between related basis states (or parts of the system) which in turn 
are governed by system-conditions. In past, this has motivated an introduction of a 
variety of random matrix ensembles as models for a wide range of complex 
systems e.g. nuclei, atoms, molecules, disordered and chaotic systems, 
quantum chromodynamics, elastomechanics, electrodynamics (see reviews 
\cite{gu,me,iz, been, alh, bohi, ve, wig, weid} and references therein for details), 
mathematical areas such as Riemann zeta function, enumeration problems in 
geometry and fluctuations in random permutations \cite{rz}, biological 
systems \cite{kwa}, stock markets \cite{pl}, atmospheric sciences 
\cite{santh} etc \cite{co1}. For example, the systems with delocalized 
wave dynamics (extended throughout the system) and antiunitary symmetries
can be well modeled by Wigner-Dyson ensembles; the latter are the Hermitian 
ensembles with Gaussian distributed matrix elements, with ratio of diagonal 
to off-diagonal variance $\alpha=2$, (originally introduced by Cartan \cite{ca},  
later developed by Wigner and Dyson to model compound nuclei and other systems) 
\cite{me,gu}. 
The cases with partially violated anti-unitary symmetries can be well-modeled by 
Dyson's Brownian ensembles (BE) \cite{dy, me} (see section VI also). 
The ensembles with arbitrary $\alpha$ ($\not=2$) \cite{rp}, banded matrices 
\cite{wig, mir1} 
(elements with non-zero variance within a band around main diagonal) and sparse 
matrices (with many elements with zero variance)  have been successfully 
used to model statistical properties of the energy levels and eigenfunctions 
of systems with localized wave dynamics (e.g. quasi one dimensional wires and 
disordered systems of higher dimensions, chiral systems) \cite{fyd, iz, mir, gg, mkc}. 
%
%
During last decade, many new ensembles were introduced to model the systems with 
unitary symmetries e.g. block form 
matrices for the cases with parity violation and pre-compound nuclei, chiral ensembles  
for systems with chiral symmetry in quantum chromodynamics \cite{ve,gu}, 
C and CI ensembles for 
cases with particle/hole structures \cite{zirn,gu}, superconductivity etc \cite{zirn}. 
The non-Hermitian 
operators e.g. scattering matrices \cite{been, me, alh}, transfer matrices \cite{been} 
or correlation matrices 
(appearing in time-series analysis e.g. stock market \cite{pl}, brain \cite{kwa}, 
atmospheric studies \cite{santh}) 
can similarly be modeled by circular ensembles \cite{me, alh}, 
Ginibre ensembles \cite{ge} and their more generic forms \cite{fz, ps0}. 
(The breadth of the subject is such that it is not possible to give a detailed account 
of all ensembles or include all references here).




The applicability of random matrix ensembles to complex systems has been 
under investigation for past few decades. The validity of the models, however, 
has been extensively verified in context of the eigenvalue fluctuation only; 
see reviews \cite{alh, me, gu, iz, be0, du, akl, ve, kwa, santh, weid} and 
references therein. 
The validity in the domain of eigenfunction fluctuations 
is so far mostly studied either in ergodic regime of the wavefunctions 
(see \cite{alh, gu, fren, fe, fal, mj, berry, psz, somm, mmms} for some original 
papers and reviews) or for quasi 1-d systems \cite{fyd} and specific cases 
e.g. disordered systems \cite{mir}. 
The growing technological demands as well as the observations of hitherto unknown 
features among eigenfunctions (e.g. multifractal structures at critical points) 
of a wide range of complex systems (for example, see \cite{ay, mir, be1, gg, kwa2}) 
have made it imperative to seek the statistical information in 
higher dimensions and beyond ergodic regime. This  motivates us to pursue 
the present study. It is also desirable to explore the 
possibilities of any connection among the critical point behavior of the eigenfunctions 
of different complex systems. One way to show the connection is by describing their various 
measures by a common mathematical formulation if possible.  
A recent study \cite{ps1,ps2,ps3}, has shown 
the existence of a similar formulation for the case of level-statistics where system 
information enters through a single parameter, basically a function of all system 
parameters. The well-known connection between the statistics of eigenfunctions  
and eigenvalues in non-ergodic regime \cite{mir} motivates us to seek a similar 
formulation for the eigenfunctions too. Such a formulation can also be 
 useful in deriving the measure of one complex system from another.



The paper is organized as follows. The section II contains a brief revision  
of the single parametric formulation of the multi-parametric probability density of 
 matrix elements for a wide range of complex systems (see \cite{ps1} for details).
The section III describes the derivation of the 
complexity parameter governed diffusion equation for the eigenfunction components 
 (of the same eigenfunction or different ones) which is used in section IV to study the 
 distribution of some of the important fluctuation measures. The other measures can also 
 be derived following the same route.  Although, the diffusion approach 
 seems to complicate the calculation by introducing a dependence on the initial 
 conditions, however, as discussed in section IV, the statistics of the system can 
 be recovered by integrating over all physically allowed initial conditions. The 
 approach has an extra advantage: it provides a common analytical base for the 
 systems which can be modeled by our ensemble (given below by eq.(1,2))  
 The section IV briefly discusses the role of complexity parameter in various 
 transitions induced due to change in system-specifics. The section V contains 
 details of the numerical verification of our analytical claims. We conclude in 
 section VI by summarizing our main results and their potential applications.

\section{Single Parametric Formulation of the Matrix Elements Probability Density}

The eigenvalues and eigenfunctions of an operator, say $H$, of a system can be 
obtained by solving the eigenvalue equation $H U_i = \lambda_i U_i$ 
(with $U_i$ and $\lambda_i$ as the eigenfunction and corresponding eigenvalue respectively)
and any other physical 
information can then be deduced, in principle, from this knowledge. In the case 
of a complex system, however, the exact form of an operator e.g Hamiltonian may not 
be known or it may be far too complicated to solve. To deal with such a situation, 
one has to make a statistical hypotheses, known as maximum entropy hypothesis, 
for $H$ \cite{ba}:  
a sufficiently complicated system can be 
described by a matrix which is as random as possible under the conditions 
compatible with the nature of the dynamics as well as the symmetry requirements. 
Thus if the symmetries and the  nature of the operation
is approximately known in a basis 
space preserving the symmetries, it can be modeled by an ensemble of full or sparse 
random matrices in that basis. For example, an  equal probability of dynamics 
in each region of a specific space suggests a uniform spread of the 
eigenfunctions  in the entire associated basis space.   The operator in such a basis 
will therefore be a full matrix, 
$\langle  k|H|l \rangle  = \sum_i \lambda_i U_{ki} U_{li}$ 
being of the same order for all combinations of basis vectors $|k \rangle , |l\rangle$ 
(with $U_{ki}$ as the $k^{th}$ component of eigenvector $U_i$). 
On the other hand, the dynamics localized in a space   
leads to variation of the eigenfunction intensities in the associated basis and the 
operator will be a sparse matrix. 




	It is clear from the above that, contrary to eigenvalues, the eigenfunction 
statistics depends on the basis in which the matrix is represented. The knowledge however 
is still relevant because (i) it can provide important information about 
the system dynamics in a given basis-space of interest, (ii) 
it is also possible to define a relevant basis to represent an operator: 
it is the basis in which the constraints on the operator appear in a natural way. 
For example, for the time-reversal invariant systems with integer angular momentum, 
the relevant basis is the one in which their Hamiltonians are simultaneously 
expressed as real-symmetric matrices \cite{bohi}.


In this paper, we consider a prototype distribution which can model a wide range of 
complex systems, namely, an ensemble of Hermitian matrices $H$, each of size $N$,  
described by a Gaussian probability density 
 
 \begin{eqnarray}
  \rho (H,h,b)=C{\rm exp}[{-\sum_{s=1}^\beta \sum_{k,l=1;k\le l}^N (1/2 h_{kl;s}). 
  (H_{kl;s}-b_{kl;s})^2 }]
 \end{eqnarray}
Here $|k \rangle , |l\rangle$ are  unit vectors of the arbitrary basis of size $N$, 
chosen to 
represent $H$ with  $H_{kl}\equiv \langle  k|H|l \rangle $. The subscript $"s"$ refers 
to the components of $H_{kl}$,  $\beta$ as their 
 total number ($\beta=1$ for real variable, $\beta=2$ for the complex one), 
 $C$ as the normalization constant, $h$  as the variance matrix with   
 $h_{kl;s}=\langle H^2_{kl;s} \rangle $ and $b$ as the mean value matrix with 
$\langle H_{kl;s} \rangle = b_{kl;s}$. Our choice of Hermitian nature of 
the ensemble restricts the present discussion to the class of systems with conservative 
dynamics. Following maximum entropy hypothesis, the above ensemble can well-describe the 
distribution of the operators for which the average behavior 
of the matrix elements and their variances is known.  Based on the complexity of the 
system, the elements of the parametric matrices $h,b$ can have various functional 
forms e.g. exponential, power law etc. For example, the limit 
 $h_{kl;s} \rightarrow 0$, corresponds to non-random nature of $H_{kl}$. 
 The limit $h_{kl;s} \rightarrow \gamma$, $b_{kl;s} \rightarrow 0$ for all 
 $\{k,l,s\}$ gives the density for a Wigner-Dyson ensemble \cite{me}: 
 $\rho(H)\propto {\rm e}^{-{\rm Tr}H^2}$. The limit 
$h_{kl;s} \rightarrow [\alpha_d \delta_{kl}+ \alpha_o (1-\delta_{kl})]$, 
$b_{kl;s} \rightarrow 0$ for all $\{k,l,s\}$ gives the density for a Rosenzweig-Porter 
ensemble \cite{rp}; (note Brownian ensembles have the same density-form too, 
see section VI). A band matrix ensemble \cite{iz, wig, mir1, fyd} with Gaussian 
distributed matrix elements and band length $t$ can be obtained by substituting 
$h_{kl;s} \rightarrow 0$, $b_{kl;s} \rightarrow 0$  if $|k-l|> t$ and 
$h_{kl;s} \propto a(|k-l|)$ for $|k-l|\le t$ with various possible forms 
of function $a$ e.g. exponential, rectangular etc.     
Similarly other ensembles with uncorrelated matrix elements, 
some of them with Gaussian randomness and others non-random, can be represented 
by appropriate choice of $h$ and $b$ parameters \cite{ps2}.       

The eq.(1) is applicable for the cases with mutually independent matrix 
elements, with no condition imposed on their moments higher than $2^{nd}$ order. 
Here we briefly mention only two such cases, namely, disordered systems and mixed 
dynamical systems; (the application to other cases e.g. algebraic or algorithmic 
complexity \cite{casa1} will be discussed elsewhere). During recent past, specific 
cases of eq.(1) have been extensively used to model 
the energy level statistics of disordered system within independent 
electron approximation; (the latter results in independent matrix elements of the 
Hamiltonian). One such example is the power law random banded matrix 
(PRBM) ensemble (each $h_{kl}$ with a power law dependence on the distance from 
the diagonal) \cite{mir1} which has been shown to be a good model for the 
level-statistics of Anderson Hamiltonian (AH) \cite{mir}. The ensemble (1) was also 
used recently to 
prove, analytically as well as numerically, the single parameter scaling of the 
level-statistics of Anderson Hamiltonian and its mapping to single parametric 
Brownian ensembles \cite{ps2}. (Later, in this paper,  AH, BE and PRBM cases are 
also used to verify our analytical predictions for the eigenfunction statistics).  

Another potential application of eq.(1) is to  systems with mixed dynamics 
where, similar to disorder, KAM tori lead to a localization of dynamics \cite{gutz}.  
The connection of quantum systems in classically chaotic and integrable 
regime to Wigner-Dyson ensembles and Poisson ensembles, respectively,  is already  
well-established \cite{be0, bohi, be2, be3}. In past, it has been suggested 
that a mixed Hamiltonian (or time-evolution operator) in a relevant basis should 
appear as a block diagonal matrix, each block being associated with an isolated 
region of the classical phase space \cite{bohi, be2, svz}. In cases, where a 
chaotic region can be decomposed in nearly, but not completely isolated subregions, 
blocks are expected to be connected through small but non-zero matrix elements. The 
average size of these matrix elements i.e the quantum constraints will be 
related to flux connecting different regions i.e to classical information.    
We further suggest that the regimes with stable islands 
can be modeled by the blocks with non-random elements (e.g. zero variance and 
non-zero mean). The chaotic regimes can be modeled by blocks with randomly 
distributed elements (e.g. same non-zero variance for all elements or only 
within a band). Based on nature of the dynamics, a chaotic block may further 
contain hierarchy of random and non-random sub-blocks; 
various diagonal blocks may also be correlated. Eq.(1) can then be applied to 
model the Hamiltonian by choosing the matrix element 
variances appropriate to the block in which they appear. 





 Eq.(1)  can not serve as a good model for the cases with correlated matrix elements. 
For example, particle-particle interactions in nuclei \cite{brody, fren2, weid, alh} 
and electron-electron interaction in disordered systems can lead to correlations 
among elements of the Hamiltonian \cite{alh, ps3}; the correlation coefficients depend 
on various system parameters. In general, such cases can occur when the interaction 
(described by $H$) between any two basis states is influenced by the other states. 
In past, consideration of particle correlations in nuclei led to 
introduction of embedded ensembles \cite{brody, fren2, gu};  however no significant 
progress has been made so far in dealing analytically with these ensembles. 

	In general, an increase of constraints on the system-dynamics 
subjects  higher moments of the matrix elements to certain specific conditions. 
This motivates us to consider an alternative ensemble, namely, 
maximum entropy ensemble with restricted higher moments. 
Within maximum entropy hypothesis, the probability 
density for such cases turns out to be non-Gaussian \cite{ps3}:
${\tilde\rho(H)}=C {\rho(H)}$ where 
\begin{eqnarray}
\rho(H) = \prod_{s=1}^{\beta}\prod_{r=1}^n {\rm exp}\left[-
\sum_{p(r)} b_{p(r)} 
\left({{\prod}_{i_p j_p}^r H_{i_p j_p;s}}\right)\right]
\end{eqnarray}
with $C$ as a normalization constant. 
Here each $H_{jk}$ is expressed in terms of its $\beta$ components, 
($\beta=1$ for the real-symmetric matrices and $\beta=2$ for the complex 
Hermitian case): 
$H_{jk} \equiv \sum_{s=1}^{\beta} (i)^{s-1} H_{jk;s}$. 
Here symbol $p(r)$ refers to a combination of $r$ matrix elements chosen 
from a set of total ${\tilde M}=N(N+1)/2$ of them; note the terms present in a
given combination need not be all different. The ${\prod^r}_{i_p j_p}$ implies a 
product over $r$ terms present in the $p^{\rm th}$ combination with 
 coefficient $b_{p(r)}$ as a measure of their correlation: 
$\langle {\prod^r}_{i_p,j_p} H_{i_p j_p;s} \rangle = 
{\partial {\rm log} C\over \partial b_{p(r)}}$. 
The $\sum_{p(r)}$ is a sum over all possible combinations (total $(\tilde M)^r$) of 
$r$ elements chosen from a set of total ${\tilde M}=N(N+1)/2$ of them.   

	The potential use of eq.(2) to disordered systems with e-e interaction 
is discussed in \cite{ps3}. Here we briefly discuss few more examples. 
The systems with chiral symmetry can be modeled by Hermitian ensembles with block 
form matrices  $H=\pmatrix{ 0 & W\cr \\ W^{+} & 0}$,  
with $W$ as a matrix of size $N$. Here, due to $H_{k',l}=H_{l,k'}^*$ for 
$1 \ge k,l \ge N$ (with $k'\equiv k+N, l'\equiv l+N$), the correlations between these 
elements are subjected to condition:
\begin{eqnarray} 
\langle H_{k',l} H_{l,k'} \rangle  &=& \langle |H_{k',l}|^2 \rangle = 
\langle |H_{l,k'}|^2 \rangle, \nonumber \\  
\langle H_{k,l} \rangle  &=& \langle H_{k',l'} \rangle
\end{eqnarray}
However, due to $H_{k,l}=H_{k',l'}=0$, all other matrix elements are uncorrelated.  
For a simple explanation, let us restrict to a case of real matrix $W$ with Gaussian 
distributed elements. The ensemble can then be represented by eq.(2) with $n=2$ 
or equivalently by density, 
\begin{eqnarray}
\rho(H,a,b) = C{\rm exp}\left[- \sum_{i\le j,k \le l}
 b_{ijkl}  H_{ij} H_{kl} - \sum_{kl} a_{kl} H_{kl} \right],
\end{eqnarray} 
with following conditions on $a$ and $b$:
\begin{eqnarray} 
b_{k',l,l,k'} &=& b_{k',l,k',l}= b_{l,k',l,k'}, \nonumber\\
a_{kl} &=& a_{k',l'}=0, \qquad 
a_{k',l} = a_{l,k'}, 
\end{eqnarray}
Note, $b$ parameters corresponding to other pairs of elements (both 
or at least one in diagonal blocks) diverge due to zero correlation between 
elements in such pairs.  
	The cases with other types of correlated blocks can similarly be 
modeled by applying  appropriate conditions on the $b$ parameters which 
correspond to combinations of matrix elements appearing in opposite blocks.   
For example, the ensemble $C$  describes the cases with particle-hole symmetry   
with a Hamiltonian $H=\pmatrix{ A & B\cr \\ B^{+} & - A^T}$ (\cite{zirn} for details).  
Now the correlations between various elements must be subjected to  
conditions $H_{k',l}=H_{l,k'}^*$, $H_{k,l}=-H_{l',k'}$; this implies another 
set of non-zero correlations (besides those given by eq.(3)): 
$\langle H_{k,l} H_{l',k'} \rangle  = - \langle |H_{k,l}|^2 \rangle = 
- \langle |H_{l',k'}|^2 \rangle$. For Gaussian distributed real matrices 
$A$ and $B$, the case can again be modeled by eq.(4) however now $b$ 
parameters for other pairs (besides those given in eq.(5)) can also be 
finite and satisfy the equality:   
$ (i) \alpha b_{k,l,l',k'} = b_{k,l,k,l}=b_{l',k',l',k'} ; 
(ii) b_{k,l,k',l} = b_{k,l,l,k'} = alpha b_{l', k',l,k'}= \alpha b_{l', k', k', l}$ 
 with $\alpha=-1$.


 Recently it was shown \cite{ps1,ps3} that the distribution $\rho$ for both cases 
(eqs.(1,2)) appear as the non-equilibrium stages of a Brownian type diffusion process 
in the  matrix-space, evolving with respect to a single parameter which is a function of 
 the distribution parameters of the ensemble:
 ...

 \begin{eqnarray}
 {\partial \rho\over\partial Y} &=& L_{+} \rho 
 \end{eqnarray}
 
 with 
 \begin{eqnarray}
 L_{\pm}= \sum_{k,l;s}{\partial \over \partial H_{kl;s}}\left[{g_{kl}\over 2}
  {\partial \over \partial H_{kl;s}} \pm \gamma H_{kl;s}\; \right]
 \end{eqnarray}
 where $g_{kl}=1+\delta_{kl}$. The variable $Y$ is the parameter governing the 
evolution of matrix elements subjected to various system conditions. For case (1), 
\begin{eqnarray} 
Y= -{1\over 2 M \gamma} {\rm ln}\left[ \prod_{k \le l}^{'}\prod_{s=1}^{\beta}
  |x_{kl;s}| \quad |b_{kl;s}|^2 \right] + C_0
\end{eqnarray}
 with $\prod'$ implying a product 
over non-zero $b_{kl;s}$ and  $x_{kl;s}=1-(2-\delta_{kl})\gamma h_{kl;s}$, 
 $C_0$ as a constant determined by the 
 initial distribution, $M$ as the number of all non-zero parameters $x_{kl;s}$ and 
 $b_{kl;s}$. The parameter $\gamma$ is arbitrary, giving the freedom to choose the 
 end of the evolution; the ${\rm lim} h_{kl;s}\rightarrow \gamma$, $b_{kl;s} \rightarrow 0$ 
 for all $k,l$ gives $Y\rightarrow \infty$ and the steady state (a Wigner-Dyson ensemble). 
 The distribution parameters being indicators of the complexity of the system, 
 $Y$ can be termed as the complexity parameter \cite{ps1}. Some 
 examples of the  calculation of $Y$ from eq.(8) are discussed in \cite{ps1}
 (for banded ensembles) and in \cite{ps2} (for Anderson Hamiltonian). The $Y$ in 
 case of a mixed system can similarly be calculated if one knows the details of 
 the mixed dynamics. 

In general, the form of parameter $Y$ for eq.(2) is quite complicated; its details can be 
found in \cite{ps3}.  However, for case (4), that is the Gaussian version of case (2),  
$Y$ can be given as 
\begin{eqnarray}
Y &=& \sum_{kl} \int {\rm d}a_{kl}\; X
+ \sum_{ijkl} \int {\rm d}b_{ijkl}\; X
+ {\rm constant }
\end{eqnarray}
where summation is implied over the distribution parameters with finite values only,   
and, $X=[\sum_{kl} f_{kl} +\sum_{ijkl} f_{ijkl}]^{-1}$ with 
$ f_{kl} = \gamma a_{kl} - 2 [\sum_{mn} a_{nm} b_{klmn} + a_{mn} b_{klnm}] $, 
$ f_{ijkl}=\left[\gamma b_{ijkl} - 2 \sum_{mn} b_{ijnm} b_{klmn} \right]$.
For further clarification we refer the reader to \cite{ps3} where an example, 
namely, the modelling of lowest Landau level of a disordered quantum 
Hall system by eq.(4) and calculation of corresponding $Y$ is discussed.

 It is easy to solve eq.(6) for arbitrary initial condition, say $H_0$ at $Y=Y_0$,: 
 $\rho(H,Y|H_0,Y_0) \propto {\rm exp}[-(\alpha/2) {\rm Tr}(H- \eta H_0)^2]$ with 
 $\alpha=\gamma (1-\eta^2)^{-1}$ and $\eta={\rm e}^{-\gamma Y}$. The probability 
 density of $H$ can now be extracted by integrating over an ensemble of initial conditions: 
 $\rho(H,Y-Y_0)= \int \rho(H,Y|H_0,Y_0)\rho(H_0,Y_0) {\rm d}H_0$. 
 It is often useful to study the statistics of the perturbed Hamiltonian $H$ in the 
 eigenfunction basis of unperturbed Hamiltonian $H_0$. Thus if the eigenfunctions of 
 $H_0$  are chosen as the basis vectors $|k \rangle ,|l \rangle$ etc, and,  
 the initial distribution is given by 
 $\rho(H_0) \propto {\rm exp}[-(1/2)\sum_j H_{0;jj}^2]$,
 the eigenvalue equation $UH=\Lambda U$ can be used to transform $\rho(H)$ from 
 matrix space to eigenvalue-eigenvector space $\{\lambda,U\}$: 
 \begin{eqnarray}
 \rho(H,Y)  \propto \prod_{k,l; k<l}^N |\lambda_k-\lambda_l|^{\beta} 
 {\rm exp}\left[-(\gamma/2) \sum_{j=1}^N \lambda_j^2 -  
 (\gamma \mu/2) \sum_{k<l} |\lambda_k-\lambda_l|^2 |U_{jk}|^2 |U_{jl}|^2 \right]
 \end{eqnarray}
 where $\mu=[{\rm e}^{2\gamma (Y-Y_0)}-1]^{-1}$. 
 
 As indicated by eqs.(6) and (10), the ensemble densities for various complex systems 
 (i.e. different $h, b$ matrices) undergo a similar evolution as a function of $Y$. 
 The $Y$-governed flow for the joint distribution of the desired eigenfunctions components 
 and eigenvalues can be obtained, in principle, by integrating either eq.(6) or eq.(10) 
over all the  undesired ones; however it is easier to integrate eq.(6). To explain the 
technique, we consider some of the important cases in this paper.

 \section{Diffusion Equation For Eigenfunction components and Related Eigenvalues}
 
 	The $k^{\rm th}$ component $U_{kl}$ of an eigenstate $U_l$ is a measure of 
the contribution of $k^{\rm th}$ basis state to the eigenstate. 
Experimental observations of complex systems indicate the level to level variations 
as well as sample to sample fluctuations of the contribution. As a result, 
the knowledge of average behavior of the components is not enough and one needs to 
study their distribution.  In this section, we consider the  joint probability 
distributions of a few relevant combinations of the components of the operator $H$. 
The basis chosen for the representation of the eigenfunctions 
is the one in which the matrix elements of $H$ have distribution (1) (or (2)). 
 We use  following notation  in reference to various correlations: 
 For a joint distribution $P_{rs}$, the subscripts $r$ and $s$ refer to the number of 
 components of each eigenvector and  the number of eigenvectors considered, 
 respectively. For example, for a joint distribution of $n$ components of $m$ 
 eigenvectors along with their eigenvalues, $r=n$ and $s=m$. 


 \subsection{Joint Distribution of a  Given Component of All Eigenfunctions 
 and Eigenvalues}

	It is often relevant to know the influence of a particular basis state  
on the system dynamics at various energies and with varying complexity 
of the system. The information can be obtained by a knowledge of the 
distribution of the same component of various eigenfunctions and its $Y$ 
governed evolution. For example, let us calculate the joint distribution 
of a given component of all eigenvectors and the eigenvalues. 
 Let $P_{1N}(Z,E,Y)$ be the probability, at a given $Y$, of finding the 
 $j^{th}$ component $U_{jn}$ of the eigenfunctions $U_n$ of $H$ between $z_{jn}$ and  
 $z_{jn}+{\rm d}z_{jn}$ and the eigenvalues $\lambda_n$ between $e_n$
 and $e_n+{\rm d}e_n$ for $n=1\rightarrow N$ (with $Z\equiv \{z_{jn}\}, E\equiv \{e_n\}$) . 
 It can be expressed as an average over entire ensemble $\rho$:
 
 \begin{eqnarray}
 P_{1N}(Z,E,Y) = \int f_N(Z,E,U,\lambda)\; \rho (H,Y) \; {\rm d}H
 \end{eqnarray}
 
 with $f_r(Z,E,U,\lambda)=\prod_{n=1}^r 
 \delta(z_{jn} - U_{jn}) \delta^{\beta-1}(z^*_{jn} - U^*_{jn}) \delta(e_n-\lambda_n)$.
 The $Y$ dependent evolution equation for $P_{1N}$ can now be derived by 
connecting the parametric derivatives of $P_{1N}$ to its derivatives 
 with respect to eigenvectors. The steps can briefly be described as follows: 
 As $Y$-dependence of $P_{1N}$ comes only through $\rho$, one can write  
 \begin{eqnarray}
 {\partial P_{1N} \over\partial Y}=\int {\rm d}H f_N\; L_+ \rho 
 = \int {\rm d}H \rho\; L_-f_N + {\tilde\gamma} P_{1N} 
 \end{eqnarray}
 with ${\tilde \gamma}=\beta N(N+1)\gamma/2$. 
 Eq.(12) is obtained, first, by differentiating eq.(11) with respect to 
 $Y$, then using eq.(6), followed by partial integration.  
 Due to $\delta$-function nature of $f_N$, its derivatives with respect to 
 matrix elements can further be reduced to the derivatives with respect to 
 $Z$ and $E$, 
 
 \begin{eqnarray}
 {\partial f_N \over \partial H_{kl;s}}
 =- \sum_{n=1}^N \left[{\partial \lambda_n \over \partial H_{kl;s}}
 {\partial f_N \over \partial e_n} 
 +{\partial U_{jn} \over \partial H_{kl;s}}
 {\partial f_N \over \partial z_{jn}}  
 +{\partial U_{jn}^* \over \partial H_{kl;s}}
 {\partial f_N \over \partial z_{jn}^*} \right] 
 \end{eqnarray}
 
 The $2^{\rm nd}$ derivative of $f_N$ can now be obtained from eq.(13) 
(see \cite{ps1}).  The substitution of eq.(13) in eq.(12) helps as the derivatives 
with respect to $z_{jn}$ and $e_n$ can be taken out of the integral. It can further 
be simplified by a  knowledge of the effect of a small perturbation of $H$ on its 
eigenvalues and eigenvectors; the related  results are given in Appendix B 
(see \cite{ps1} for the details). Using the relations, eq.(12) can be rewritten as 

 \begin{eqnarray}
 {\partial P \over\partial Y}= \left( L_Z + L_Z^* \right) P + L_E P
 \end{eqnarray}
 where $P=C_1 P_{1N}$, $C_1={\rm e}^{-{\tilde\gamma} Y}$, and 

 \begin{eqnarray}
 L_Z &=& {\beta^2 \over 4} \sum_{n,m=1;n\not=m}^N 
 {1\over (e_n-e_m)^2} {\partial \over \partial z_{jn}}
 \left[ {\partial \over \partial z^*_{jn}} |z_{jm}|^2 
 -  {\partial \over \partial z_{jm}} z_{jn} z_{jm}  
 +  z_{jn}\right], \nonumber \\
 L_E &=& 
 \sum_{n} {\partial \over \partial e_n}\left[
   \gamma e_n  + \sum_{m;m\not=n} {\beta \over e_m-e_n}
 + {\partial \over \partial e_n} \right]. 
 \end{eqnarray}
 
 where $L_Z^*$ implies the complex conjugate of $L_Z$; note $L_Z=L_Z^*$ 
for $\beta=1$ case. Eq.(14) describes the $Y$-governed diffusion of a given 
component of all eigenvectors  and all eigenvalues. 
Its solution depends on the choice of initial condition $H_0$. In 
the diagonal representation of 
$H_0$ (taking $\rho(H_0) \propto {\rm e}^{-(1/2)\sum_j H_{0;jj}^2}$, the solution 
can be given as
\begin{eqnarray}
 P_{1N} \propto \prod_{k,l; k<l}^N |e_k-e_l|^{\beta} 
 {\rm exp}\left[-(1/2) \sum_{j=1}^N e_j^2 - 
 (\mu/2) \sum_{m<n} |e_n-e_m|^2 |U_{jn}|^2 |U_{jm}|^2 \right]
 \end{eqnarray}
 with $\mu$ same as in eq.(10). 
(Note, the above result can directly be obtained from eq.(10) too).

 \subsection{Joint Distribution of all Components of A given Eigenfunction and 
its Eigenvalue}
 
The distribution of the components of a 
specific eigenstate contains information about various basis states contributing 
to the state which in turn determines its spread.   
 Proceeding along the same lines as for $P_{1N}$, the diffusion equation for the 
joint probability 
 $P_{N1}$ of the components $U_{nk}$, $n=1 \rightarrow N$, of an eigenvector $U_k$ and 
 the corresponding eigenvalue $\lambda_k$ can also be obtained. The evolution of 
 
 \begin{eqnarray}
 P_{N1}(Z_k,e_k,Y)= \int\; {\tilde f_k} \; \rho (H,Y)\;  {\rm d}H, 
 \end{eqnarray}
 
 with $\tilde f_k= \delta(Z_k - U_k) \delta^{\beta-1}(Z_k^* - U_k^*) 
\delta(e_k-\lambda_k)$,  can again be shown to be described by 

\begin{eqnarray}
{\partial P_{N1} \over\partial Y}= F_k+F_k^*+L_{e_k}P_{N1} 
\end{eqnarray}
where $F_k = (\beta^2/4) \sum_{q=1}^2 L_{qk} $
with 
\begin{eqnarray}
 L_{1k}  &=& \sum_{n=1}^N{\partial \over \partial z_{nk}}
 \left[  z_{nk} Q_{nn;k}^{02} \right] \nonumber \\
 L_{2k} &=&   
 \sum_{m,n=1}^N{\partial^2 \over \partial z_{nk} \partial z^*_{mk}}
 Q_{mn;k}^{12}, \nonumber \\
 L_{e_k}P_{N1} &=& {\partial \over \partial e_k}\left[
 \gamma e_k P + {\partial P \over \partial e_k} + \beta Q_{nn;k}^{01}\right]
\end{eqnarray}
and
\begin{eqnarray}
Q_{mn;k}^{rs} &=& \sum_{j; j \not=k} \int {(z_{nj} z^*_{mj})^r \over (e_k-e_j)^{s}} \;
{P_{N2}} \;  {\rm d}\tau_j. 
\end{eqnarray}
Here ${\rm d}\tau_j \equiv {\rm d}e_j {\rm d}^{\beta}Z_j$ with 
${\rm d}^{\beta} Z \equiv {\rm d}Z {\rm d}Z^*$ 
and $P_{N2}=P_{N2}(Z_k,Z_j,e_k,e_j)$ is the joint probability of all components of 
the two  eigenvectors $Z_j\equiv \{z_{nj}\}$ and $Z_k\equiv\{z_{nk}\}$ 
($n=1\rightarrow N$) alongwith  their eigenvalues $e_j$ and $e_k$, respectively: 
 \begin{eqnarray}
 P_{N2}=\int\; {\tilde f_k}\;{\tilde f_j}\;\rho (H,Y)\;  {\rm d}H. 
 \end{eqnarray}

 The presence of eigenvalue-eigenfunction correlations in exponent of $\rho(H)$ 
(e.g. eq.(10)) as well as the terms of type $(e_j-e_k)^{-2}$ in the denominator 
of eq.(20)  makes it difficult to write $F_k$ (in eq.(18)) as a function of 
$P_{N1}(Z_k,e_k)$. To write eq.(18) in a closed form, it is necessary to 
approximate $Q_{mn;k}^{rs}$ (Appendix A): 

\begin{eqnarray}
Q_{mn;k}^{rs} &\approx & D_k^{-s}(N-1)^{1-r}\chi^{s/2} 
[\delta_{mn}-z^*_{mk}z_{nk}]^r P_{N1} 
\end{eqnarray} 
with $D_k$ as the local mean level spacing at energy $e_k$. Here
$\chi =1$ for $\mu < \zeta_k^d$, 
$\chi \sim  (\mu/\zeta_k^d)$ for $\mu > \zeta_k^d$ 
with $\zeta$ as the ensemble averaged  localization length of the eigenfunction 
$U_k$ and $d$ as the system-dimension. 
The length $\zeta$ enters in the formulation due to its relation with typical intensity of 
a wavefunction: $|U_{nk}|^2 \equiv |z_{nk}|^2 \sim  \zeta_k^{-d}$

 The substitution of eq.(22) in eq.(19) helps to write $F_k$ in terms of  
$P_{N1}$, thus  reducing the evolution equation (18) for $P_{N1}$ in a closed form:  
 
\begin{eqnarray}
F_k &=& {\beta^2 \over 4 D^2} \sum_{n=1}^N   {\partial \over \partial z_{nk}}
\left[ \sum_m {\partial h_2\over \partial z^*_{mk}} 
+  h_1  \right]  
 \end{eqnarray}
with $h_1= (N-1) \chi z_{nk} P_{N1}$, $h_2=\chi (\delta_{mn}- z_{nk} z^*_{mk}) P_{N1}$.

 \subsection{Joint Distribution of all components of $q$ Eigenfunctions and 
their eigenvalues}
 
 For certain physical properties e.g susceptibility,  a knowledge of the correlations 
among  two (or more)  
 eigenvectors at two different space points may be required. The fluctuations of such 
 correlations can be determined by the joint probability density 
 $P_{Nq}$ of the components $U_{nk}$ ($n=1 \rightarrow N$) of $q$ eigenvectors $U_k$
 ($k=1 \rightarrow q$) where 
 
 \begin{eqnarray}
 P_{Nq}(Z_1,Z_2,..Z_q,Y)= \int\; \prod_{k=1}^q {\tilde f_k}\; \rho (H,Y)\;  {\rm d}H, 
 \end{eqnarray}
 
 Proceeding exactly as in previous two cases, the $Y$-governed diffusion of $P_{Nq}$ 
can be shown to be described as   
    
 \begin{eqnarray}
 {\partial P_{Nq}\over \partial Y} &=&  \sum_{k=1}^q 
\left[ {\tilde F}_k + {\tilde F}_k^* +L_{e_k}P_{Nq}  \right] 
\end{eqnarray}
where 
\begin{eqnarray}
{\tilde F}_k &=& F_k + {\beta^2 \over 4} \sum_{l=1;\not= k}^q\sum_{m,n=1}^N 
{\partial^2 \over \partial z_{nk} \partial z_{ml}} 
\left[ {z_{nk} z_{ml} \over (e_k-e_l)^2}\right] P_{Nq}. 
\end{eqnarray}

Note although $F_k, L_{1k}, L_{2k}$ are still defined as in eqs.(18, 19) however the 
definition of $Q$ is now slightly altered with $P_{N(q+1)}$ replacing $P_{N2}$ 
in eq.(20).  
Here $P_{N(q+1)}$ is the joint probability density of $q+1$ 
eigenfunctions, namely,  
$Z_1,Z_2,..,Z_q$ alongwith $Z_j$ (with $j>q$). 
Similar to previous case, the integral  $Q$  can again be approximated 
so as to express $F_k$ in terms of $P_{Nq}$:
$ Q_{mn;k}^{rs} \approx  D_k^{-s}(N-1)^{1-r}\chi^{s/2} 
[\delta_{mn}-\sum_{l=1}^q z^*_{ml}z_{nl}]^r P_{Nq}$. 
 Here again $\chi=1$ for $\mu < \zeta_k^d $ and 
$\chi \sim (\mu/\zeta^d)$ for $\mu > \zeta_k^d$.

The above approximation for $Q$ leaves the expression for $F_k$ in  the same form  
as in eq.(23) however now $h_1=\chi (N-1) P_{Nq}$, and, 
$h_2=\chi (\delta_{mn}-\sum_{l=1}^q z_{ml}^* z_{nl}) P_{Nq}$. The substitution of $F_k$ 
in $\tilde F_k$ gives the latter as a function of $P_{Nq}$ which in turn 
 reduces eq.(26) in a closed form for $P_{Nq}$. The equation can then be used, 
by integrating over undesired components, to obtain the distributions of various 
combinations of eigenfunction components.

 \section{Diffusion Equation For Fluctuation Measures of Eigenfunctions}
 
 The ensemble average of any measure of the eigenfunction correlations  can be expressed 
 in terms of $P$ ( $P\propto P_{rq}$ for a correlation function of 
 $r$ components of $q$ eigenstates). For example, the average of 
 a measure, say $C$, 
 describing the correlation among a set $X$ of eigenfunction components can be written as  
 \begin{eqnarray}
 \langle C(X;Y) \rangle =\int_0^{\infty}  C(X;Y) P(X;Y) {\rm d}X.  
 \end{eqnarray}  
where $\langle . \rangle $ denotes an averaging over various realizations of the sample.
However the strength of the reproducible fluctuations of the correlations in 
different realizations of  same complex system is of the order of the averages. 
As a consequence, a knowledge of just the averages is not enough and it is necessary 
to know the distributions of correlations.  

 The $Y$-governed evolution of the distribution $P_C$ of a measure $C$    
 can be obtained by an integration of the undesired variables in eq.(14) 
 (or eq.(18), eq.(25) as per requirement).  
 As examples, we derive the evolution equations for few important measures in this 
 section. The involved integrals are, however, quite tedious and analytical 
 approximations seem necessary to reduce the equation in a closed form. 
 As a check on our results, we study the $Y\rightarrow \infty$ limit of each 
 measure. This limit corresponds to the flow of ensemble (1) (and ensemble (2)) 
 to its steady state, that is, a Wigner-Dyson Ensemble. As a consequence, each 
 measure is expected to evolve to  its Wigner-Dyson limit as $Y\rightarrow \infty$. 
 We verify our results numerically too; the details are given in section VI.

 \subsection{Distribution of Local Eigenfunction Intensity}

The distribution function of the local eigenfunction intensity i.e. the eigenfunction 
intensity $u$ at a given basis state, say $n$ can be defined as 
$P_u(u,e)=\langle  \sum_{k=1}^N \delta(u- N |z_{nk}|^2) \delta(e-e_k)\rangle $. 
The diffusion of $P_u$ as a function of $\Lambda$ can be obtained from either eq.(14) 
with $P\propto P_{1N}$ or eq.(18). For technical simplification, 
however, we choose the former and first study the evolution of the distribution 
$P_{11}(x,e)$  of an eigenfunction component $x=N^{1/2} z_{nk} =(u^{1/2})$ at an 
energy $e$, defined as  
 \begin{eqnarray}   
 P_{11} (x,x^*,e)= \langle   \delta^{\beta}_x \delta_e \rangle 
= \int \delta^{\beta}_x \delta_e  P_{1N}(Z,E,Y) \; 
 {\rm d}E \; {\rm d}^{\beta}Z
\end{eqnarray}
 where $\delta^{\beta}_x= \delta(x- \sqrt{N} z_{nk}) 
\delta^{\beta-1}(x^*- \sqrt{N} z^*_{nk})$ and $\delta_e= \delta(e-e_k)$ 
and ${\rm d}\tau \equiv {\rm d}E {\rm d}^{\beta}Z$. 
 The diffusion equation for $P_{11}(x,e)$ can be obtained by integrating eq.(14), with  
 $P \propto P_{1N}$, over the variables $e_j$ and $z_{nj}$, $j=1 \rightarrow N$,

 \begin{eqnarray}
 {\partial P_{11} \over\partial Y}=
{\beta^2 \over 4 }\left[ 2{\partial^2 G_1 \over\partial x \partial x*}
 +  {\partial (x G_0) \over\partial x} + 
    {\partial (x^* G_0) \over\partial x^*}\right]
 + L_e P_{11}
 \end{eqnarray}
 where 
 
 \begin{eqnarray}
 G_r(x,e) \equiv   \sum_{j;j\not=k}\int \delta^{\beta}_x 
 \delta_e {|z_{nj}|^{2r} \over (e_k-e_j)^{2} } 
 P_{1N} \; {\rm d}\tau, 
 \end{eqnarray}
with $r=0,1$ and 
$\int \delta^{\beta}_x\; \delta_e\; [L_E P_{1N}]\; {\rm d}E\; {\rm d^{\beta}}Z = L_e P_{11}$.

Eq.(29) describes the sensitivity of the local intensity distribution to the energy 
scale  $e$ as well as various system parameters. As discussed in appendix A 
(see eq.(A7)), $G_r$ can be approximated as

\begin{eqnarray}
G_r &\approx &  \mu \chi_0 (N-1)^{1-r}(N-|x|^2)^r P_{11}(x)/D_k^2  
\qquad\qquad  
\end{eqnarray} 

with $\chi_0=\mu^{-1}$ for $\mu |x|^2 < 1$ and 
$\chi_0 \sim  |x|^2$ for $\mu |x|^2 > 1$ where  
$\mu=[{\rm e}^{2\gamma (Y-Y_0)}-1]^{-1}$ and $D_k$ as the local mean level 
spacing at energy $e_k$.
A substitution of approximated $G_r$ in eq.(29) and an integration 
over $e$ gives the energy-averaged local intensity distribution
$P_x(x)=\int P_{11}(x,e){\rm d}e$:
 
 \begin{eqnarray}
 {\partial P_x \over\partial \Lambda_u}={\beta^2 \over 4} \left[
 2  {\partial^2 [h_2(x)  P_x] \over\partial  x \partial x^*}
 +  {\partial [h_1(x)  P_x] \over\partial {x}}
 +  {\partial [h_1(x^*) P_x] \over\partial {x^*}}\right]
 \end{eqnarray}
with $h_2(x)=\chi_0 (N-x^2)$, $h_1(x)=\chi_0 (N-1) x$.
Here $\Lambda_u={\mu \Lambda}$ with $\Lambda=(Y-Y_0)/ D_k^2$. 
Eq.(32) suggests that the evolution of $P_x$ is governed by 
 a rescaled parameter $\Lambda_u$ instead of $Y$.

 For cases $|x|^2 << N $ (thus $\Lambda_u=\Lambda$), the above equation can easily be solved: 
 $P_x(x,\Lambda|x_0) \propto {\rm e}^{-\beta |x-\gamma x_0|^2/2(1-\gamma^2)}$ 
 with $\gamma={\rm e}^{-\beta N\Lambda/2}$ and $P_{x_0}(x_0)$ as the initial 
 distribution. The steady state limit 
${\partial P_x \over\partial \Lambda} \rightarrow 0$ 
of eq.(32) occurs at $\Lambda \rightarrow \infty$. The solution in this limit 
corresponds to Wigner-Dyson case i.e.  
$P_x(x, \Lambda \rightarrow \infty) \propto {\rm e}^{- \beta |x|^2/2}$ 
 or, equivalently, Porter-Thomas distribution  
 $P_u(u,\Lambda\rightarrow \infty) \propto u^{(\beta-2)/2} {\rm e}^{-\beta u/2}$ 
\cite{alh, me}(using $u=|x|^2$, which gives  $P_u=P_x (2|x|)^{-1}$). 
 
It is desirable to know the solution $P_x$ of eq.(32), or alternatively, 
$P_u$  for finite, non-zero $\Lambda_u$ and 
all ranges of $u$. In the diagonal representation of $H_0$, which corresponds 
to an initial distribution $P_{u_0}(u_0,\Lambda=0)=N^{-1} [\delta(u-1)+(N-1)\delta(u)]$,
eq.(32) gives following short range behavior of  $P_u$:  
 
\begin{eqnarray}
P_u &=&(\beta u/2)^{\beta/2-1} 
{{\rm e}^{-\beta u/2}\over \Gamma(\beta/2)}\left[1+{\kappa\over 2}
\left((\beta+2)/\beta-(\beta+2){\sqrt u}+\beta u/2\right)+.. \right] 
\qquad {u \lesssim \kappa^{-1/2}} \\
& \approx & (\beta u/2)^{\beta/2-1} {1\over \Gamma(\beta/2)} 
{\rm exp} \left[(\beta/2)\left (-u +{\kappa\over 2} u^2+...\right) \right]   
\qquad {\kappa^{-1/2} \lesssim u  \lesssim \kappa^{-1}} 
\end{eqnarray}
 where $\kappa={\rm e}^{-2\beta N \Lambda}$ (note $\kappa \approx \mu$ in large 
$Y$-limit and for $D_k^2 \sim (\beta N)^{-1}$) . 

The tail behavior of a distribution has a significant influence on its moments 
and the related physical properties.  The asymptotic analysis of eq.(32) shows 
$P_u(u)$ to be a broad distribution:
 
\begin{eqnarray}
 P_u(u) & \simeq & {\rm exp}\left[-\alpha_{u0} {u^{1/2}} -  
 \sum_{n=1}^{M} \alpha_{un} {\rm ln}^n (\kappa u)  \right]  
 \qquad {u \gtrsim \kappa^{-1}} 
 \end{eqnarray}
Here the coefficients  are sensitive to system-specifics:
$\alpha_{u0} \simeq 4 q_1 \beta^{-1}({\rm e}^{\beta N\Lambda_u}-1)$, 
$\alpha_{u1} \simeq -N/4,  \alpha_{u2} \simeq (N\beta/16){\rm e}^{\beta N\Lambda_u} $, 
$\alpha_{un;n>2}\simeq (-1)^n (\nu_{n} \beta^2 N/4){\rm e}^{2\beta N\Lambda_u}$ 
with $\nu_n$ decreasing as $n$ increases. The decreasing coefficients alongwith alternate 
$\pm$ signs lead to near-cancellation of higher order terms (with $n>2$) in the 
exponent.  Consequently, the tail is dominated by a log-normal behavior for the systems 
with large, finite $\Lambda$-strengths  and a weaker than exponential decay in 
$\Lambda \rightarrow 0$ limit. 


 Eq.(35) indicates the existence of a log-normal  asymptotic tail for the local eigenfunction 
intensity of any complex system with finite, non-zero $\Lambda_u$. A log-normal behavior 
of $P(u)$ suggests a power-law behavior of its moments: 
$\langle u^q \rangle  \propto N^{-d_q}$ \cite{mj}.
Here $d_q$ is an effective dimension which can be different from a spatial 
dimension $d$. The form of $P_u(u)$ at finite $\Lambda$ is therefore fixed by a spectrum 
of scaling exponents (as the moments can be used to recreate the distribution); 
the situation is termed as multifractal scaling. 
            Further, as shown later, a log-normal tail of P(u) results in the similar 
behavior of the distributions of other related correlations and physical properties. 
Such a behavior has already been indicated for the physical properties e.g. conductance, 
density of states, local density of states and relaxation time etc. of disordered 
systems \cite{mir}.

	The significance of above $P(u)$-formulation is that here system dependence 
 (other than size) enters only through one parameter, namely, $\Lambda$. This being valid 
for any complex system, modeled by eq.(1) (and eq.(2)), is thus applicable for disordered 
systems too. It is therefore relevant to compare our result with those obtained for 
disordered systems using other techniques (using renormalization group theory approach for 
dimension $d=2+\epsilon$, $\epsilon<1$ \cite{akl}, and, by using Berezinski and 
Abrikosov-Ryzkhin techniques for strictly $d=1$ cases \cite{ap, me, aa}; the techniques 
predict a ${\rm e}^{-\alpha_1 u^{1/2}}$ tail for $d=1$, a log-normal tail for $d=2$ 
and a log-cube tail for $d=3$ case. However our technique predicts a  log-polynomial 
behavior however dominated by log-normal term for all dimensions.

 \subsection{Inverse Participation Ratio (IPR)}  
 
  The $q^{th}$ order inverse participation ratio $I_q$ of an eigenvector, 
say $U_k$, is defined as  $I_q(k)=\sum_{j=1}^N |U_{jk}|^{2q}$. The physical meaning 
of $I_q$ can be illustrated by two limiting cases: (i) an eigenfunction with 
identical components $U_{jk}=N^{-1/2}$ corresponds to $I_q(k)=N^{1-q}$, and, 
(ii) an eigenfunction with only one non-zero component (say $n^{\rm th}$) which 
gives $U_{jk}=\delta_{nk}$ and $I_q(k)=1$. 
The case (i) corresponds to completely ergodic eigenfunctions covering 
 randomly but uniformly the whole sample of volume $V$. The case (ii) corresponds 
to a wavefunction localized in the neighborhood of a single basis state. 
Thus $I_q$, in general, is related to  reciprocal of the number of components 
significantly different from zero and contains information about spread of a 
wavefunction in the basis space. For example, for a d-dimensional exponentially 
localized state, $I_2 \sim (a/\zeta)^d$, where $a$ and $\zeta$ are the lattice 
constant and localization length, respectively. Consequently, the typical value 
of $I_2$ is a frequently used characteristic of the eigenfunction 
localization \cite{mir}:
 $I_2^{typ}={\rm exp}{\langle {\rm ln}I_2 \rangle }\approx N^{-D_2}$ with 
 $D_2$ a system dependent scaling exponent (also known as correlation dimension).

  The ensemble average of $I_q$ is related to $q^{\rm th}$ moment of the 
distribution $P_x (x)$: 
$\langle  I_q \rangle =N^{1-q} \int_0^{\infty} |x|^{2q} P_x (x) {\rm d}^{\beta} x = 
\int I_q P_{I_q} \; {\rm d}I_q $, 
The average inverse participation ratios can therefore provide 
information about the scaling exponents. As a consequence, it is useful to know  
the effect of changing system parameters on $\langle  I_q \rangle $. 
Due to $P(u)$ decay for the ranges $\mu u \gtrsim 1$, major contribution to  
$\langle  I_q \rangle $ comes from the region $\mu u \le 1$.  From eq.(32), 
it can be shown that 
 
 \begin{eqnarray}
{\partial \langle  I_q \rangle  \over\partial \Lambda} 
\approx q \alpha \langle  I_{q-1}\rangle - q t \langle I_q \rangle. 
 \end{eqnarray}
where $\alpha=2q+\beta-2$, $t=N\beta+2q-2$. Eq.(36) depends on two parameters, 
namely $\Lambda$ and $t$ which results in a different power law behavior for 
each $\langle  I_q \rangle $,  

\begin{eqnarray}
\langle  I_q(\Lambda)\rangle= {\rm e}^{-q t\Lambda}\left[\langle I_q(0)\rangle + 
\alpha \int_0^{\Lambda} \langle  I_{q-1}(r) \rangle 
{\rm e}^{qtr} {\rm d}r \right]. 
\end{eqnarray}
For $\Lambda \rightarrow \infty$, eq.(37)  gives a correct steady state limit, 
namely, Wigner-Dyson behavior: $\langle  I_q \rangle  \rightarrow 
{\alpha\over t} \langle I_{q-1} \rangle $ or 
 $\langle  I_q \rangle  ={(2q)!\over 2^q q!} N^{1-q}$ for $\beta=1$ and 
$\langle  I_q \rangle =q! N^{1-q}$ for $\beta=2$.  
For finite nonzero $\Lambda$, $\langle  I_q \rangle $ can be determined if 
$\langle  I_{q-1}(\Lambda) \rangle $ as well as some past information about 
the system (to choose it as an initial state which will give $ \langle  I_q(0) \rangle$) 
is known. For example, for 
the systems where completely localized wavefunction dynamics is a valid physical 
possibility (e.g. disordered systems, mixed systems etc.), it can be chosen as the 
initial state which corresponds to $\langle  I_q(0) \rangle =1$; this gives 
$\langle  I_1(\Lambda)\rangle=1$, $\langle I_q(\Lambda) \rangle 
\approx {\rm e}^{-q \beta N \Lambda}$  for $q <N$.

	In general, the IPR fluctuations reflect the level to level variations of 
the spatial structure of eigenfunctions. In a complex system e.g. nano-system, 
however,  the sample to sample fluctuations of the eigenfunctions also  manifest 
themselves through IPR fluctuations which makes a knowledge of the $I_q$-distribution 
over whole ensemble of samples relevant. The distribution $P_{I_q}$ of $I_q$ of 
an eigenfunction, say $Z_k$, with the 
 components $\{z_{nk}\}_{k=1,..N}$ is related to  $P \propto P_{N1}$: 
${\bf P_{I_q}(I_q)= \int \delta_{I_q} P_{N1}(Z_k,e_k,Y) \; {\rm d}e_k \; {\rm d}\tau_k}$ 
with $\delta_{I_q} \equiv \delta(I_q-\sum_n |z_{nk}|^{2q})$ and the volume element 
${\rm d}\tau_k$ same as in eq.(20).  
The $Y$ governed evolution of $P_{I_q}$ can therefore be obtained from Eq.(18) for 
 $P \propto P_{N1}$: 
 
 \begin{eqnarray}
 {\partial P_{I_q} \over\partial Y}= {\beta^2\over 4} (X_1  + X_2) + X_3
 \end{eqnarray}
 where $X_3= \int \delta_{I_q} [L_E P_{N1}] {\rm d}\tau_k  = 0$ and 
 \begin{eqnarray}
 X_1 &=&   \int \delta_{I_q} \left[ L_{1k} + L_{1k}^* \right] 
 {\rm d}e_k {\rm d^{\beta}}Z_k   \\
 &=& {4\over \beta} {\partial \over \partial I_q}  I_q
   \int \delta_{I_q} F_1 \; {\rm d}\tau_k, \\
 X_2 &=& 2 \int \delta_{I_q} L_{2k}  {\rm d}\tau_k   \\
 &=& {8 q^2\over \beta^2} {\partial^2 \over \partial I_q^2} \int \delta_{I_q}
 F_2 \; {\rm d}\tau_k   
 - {4 q (2q+\beta-2)\over \beta^2} {\partial \over \partial I_q} \int \delta_{I_q}
 F_3 \; {\rm d}\tau_k 
 \end{eqnarray}
 with $F_1=\left[ Q^{02}_{nn;k} + Q^{02 *}_{nn;k} \right]$, 
$F_2=\sum_{m,n} |z_{mk}|^{2(q-1)} |z_{nk}|^{2(q-1)} z_{nk}^* z_{mk}Q^{12}_{mn;k}$   
 and $F_3=\sum_{n} |z_{nk}|^{2(q-1)}  Q^{12}_{nn;k}$ 
where $L_{1k}, L_{2k}$ and $Q_{mn;k}$ are given by eqs.(19, 20).
Using the approximate form (22) for $Q^{rs}_{mn;k}$, $F$'s  can further be reduced:
\begin{eqnarray}
F_1 &\approx& q \chi (N-1)/D^2 \nonumber \\
F_2 & \approx &  {\chi \over D^2} \left[ \sum_n |z_{nk}|^{2(2q-1)} - 
 \left(\sum_n |z_{nk}|^{2q} \right)^2 \right]  P_{N1} \nonumber \\
F_3 & \approx & {\chi \over D^2} \sum_n \left[|z_{nk}|^{2q-2}-|z_{nk}|^{2q}\right]P_{N1}. 
 \end{eqnarray}
 where $\chi=1$ for $\mu < \zeta_k^d$ and $\chi \sim \mu/\zeta_k^d$ for 
$\mu  > \zeta_k^d$.

In general, the fluctuations of different moments (or measures) of the eigenfunction 
intensity need not be mutually independent. We can therefore define the joint 
distribution of two measures, say, $h_1(z),h_2(z)$:   
\begin{eqnarray}
P_{h_1,h_2}(h_1, h_2)=\int \delta[h_1-h_1(z)] \delta[ h_2-h_2(z)] 
P_{N1} {\rm d}e_k {\rm d}^{\beta} Z_k
\end{eqnarray}
The above definition along with the equality
$\sum_n |z_{nk}|^{2(2q-1)} = \sum_{m,n} |z_{mk}|^{2q} |z_{nk}|^{2(q-1)}
-\sum_{m,n;m\not=n} |z_{mk}|^{2q} |z_{nk}|^{2(q-1)} $ gives

\begin{eqnarray}
\int \delta_{I_q} \left[\sum_n |z_{nk}|^{2(2q-1)}\right] P_{N1} {\rm d}e_k {\rm d}^{\beta} Z_k 
&\approx & 
\int I_q I_{q-1} P_{I_{q,q-1}}(I_q,I_{q-1}) {\rm d}I_{q-1} 
- \int  W  P_{I_q,W}(I_q,W) {\rm d}W \nonumber \\ 
&\approx &   I_{q-1}^{typ} I_q \; P_{I_q} - W_q^{typ}  P_{I_q}
\end{eqnarray}
where $W_q$ is a measure of the  correlation between 
the intensities localized at two different basis sates:
$W_q=\sum_{m,n;m\not=n} |z_{mk}|^{2q} |z_{nk}|^{2(q-1)} $. The 
$2^{nd}$ equality in eq.(45) is obtained from first by replacing 
$I_{q-1}$ and $W$ by their typical values; the superscript $typ$ 
over a variable $R$ indicates its typical value: 
$R^{typ}={\rm exp}\langle {\rm ln}R \rangle$.)
Using eqs.(44,45), the terms $X_1$ and $X_2$ can be rewritten as 
the functions of  $I_q$ and $P_{I_q}$ which in turn leads to

 \begin{eqnarray}
   {\partial P_{I_q} \over\partial \Lambda_{ip}} \approx 
 2 q  {\partial^2  \over\partial {I_q}^2} 
\left[ I_{q-1}^{typ} I_q-  W_q^{typ}  - I^2_q \right] P_{I_q}
 - {\partial \over\partial I_q} \left[\alpha  I_{q-1}^{typ}  - t I_q  \right] P_{I_q}
 \end{eqnarray}
 with $\alpha, t$ same as in eq.(36) and 
\begin{eqnarray}
\Lambda_{ip}= q \chi \Lambda.
\end{eqnarray}
Note, the above equation alongwith the definition $\langle  I_q \rangle = 
\int I_q P_{I_q} \; {\rm d}I_q $ again leads to eq.(36).

	The behavior of $P_{I_q}$ in different $I_q$ regimes can now be probed 
by analyzing eq.(46), using completely localized eigenstates as the initial state. 
The behavior varies from an exponential decay for small-$I_q$ regime to 
log-power law decay for asymptotic tail regime of $I_q$:
 
\begin{eqnarray}
 P_{I_q} & \simeq &  {\rm exp}\left[-\alpha_{i0} I_q \right]  \qquad {I_q \lesssim 
  e^{-1} I_q^{typ} } \\ 
  & \simeq & { I_q^{-1}} {\rm exp}\left[ - \sum_{n=1}^{M} \alpha_{in} {\rm ln}^n 
( I_q/ I_q^{typ} ) \right]  \qquad { I_q  \gtrsim e^{-1}  I_q^{typ}  } 
\end{eqnarray}
with $e\approx 2.72$, $\alpha_{i0} \approx t (1-{\rm e}^{-t\Lambda_{ip}})^{-1}/2q$ and  
$\alpha_{in} \sim (-1)^n 2 ( I_{q-1}^{typ})^n /(2q I_q^{typ})^n $ 
(valid for $q<N$) and $M$ as a large integer. 
Note the alternate $\pm$ signs of terms with increasing powers lead to convergence 
of the series in the exponent.  However the tail is dominated by increasingly higher  
powers of the logarithmic term as $I_q$ increases above its typical value. For example, 
for $e^{-1} I_q^{typ} < I_q \lesssim I_q^{typ}$, $n=1$ dominates the exponent and  
$P_{I_q}$ behaves as a power-law. Similarly, the tail shows a log-normal decay for 
regime $I_q^{typ} < I_q \lesssim e I_q^{typ}$.

  Eq.(46) depends on more than one parameter, namely, $\Lambda_{ip}$ as well as 
size-dependent parameters (appearing through $t$). This suggests an absence of 
single parameter scaling in IPR distributions. However, as suggested by eqs.(48, 49), 
it seems possible to define a single parameter locally (that is, different 
single parameters governing different IPR regimes). Further note that the 
asymptotic behavior of $P_{I_q}$ is sensitive to $\Lambda$-strength and is therefore 
system-specific. This result also agrees with the NLSM-result obtained for 
disordered systems \cite{mj}. 

  .

 \subsection {Pair-Function w(r,r')}
 
  The measure contains important information about the 
 spatial correlations between components of an eigenfunction $Z_j$ at two different 
 basis points of the sample and at an energy $e$: 
 $w(n,m)= |z_{nj} z_{mj}|^2$; (equivalently $w(r,r')=|z_j(r) z_j(r')|^2$ in a 
continuous basis e.g. coordinate space $r$).   In the localized phase, the asymptotic 
 behavior of ${\rm ln} w(r,r')$ at $|r-r')|\rightarrow \infty$ determines the 
 rate of exponential decay of the eigenfunction amplitude. It is also 
 useful for many physical applications e.g. in determination of the form factor 
 of resonance scattering in the complex nuclei or the resonance conductance of 
 the quantum dot with point contacts in the coulomb blockade regime \cite{pg}.

         The distribution 
$P_{w,e}=\langle  \sum_j\delta(w-|z_{nj} z_{mj}|^2)\delta(e-e_j)\rangle$                   
of the correlation between $n^{\rm th}$ and $m^{\rm th}$ components  of an 
eigenfunction, at an energy $e$,  is related to $P_{N1}$: 
${\bf P_{w}(w,e)= \sum_j \int \delta_w \delta_{e,j} P_{N1}(Z_j,e_j,Y) \;  
{\rm d}\tau_j}$
with $\delta_w \equiv \delta(w-|z_{nj} z_{mj}|^2)$ and 
${\rm d}\tau_j$ same as in eq.(20). Consequently, its rate of change 
with respect to $Y$ can be determined by eq.(18),  
 
 \begin{eqnarray}
 {\partial P_{w} \over\partial Y}= N L_e P_w + {\beta^2\over 4} (A_1+A_2)
 \end{eqnarray}
 where
\begin{eqnarray}
 A_1  &=& {4\over \beta} \sum_j  {\partial \over \partial w} 
   w \int \delta_w \delta_{e,j} \left[ Q^{02}_{nn;j} +Q^{02}_{mm;j} \right] \; 
 {\rm d}\tau_j   \\
 A_2 &=& {8\over \beta^2} \sum_j {\partial^2 \over \partial w^2} w 
\int \delta_w \delta_{e,j}
 \left[ F_1+F_2 \right] \; {\rm d}e_j {\rm d^{\beta}}Z_j   
 - {4\over \beta} \sum_j {\partial \over \partial w} \int \delta_w \delta_{e,j}
 \left[ F_1+ 2\beta^{-1} F_2  \right] \;  {\rm d}\tau_j 
 \end{eqnarray}
 with 
 $F_1=|z_{mj}|^2 Q^{12}_{nn;j}+ |z_{nj}|^2 Q^{12}_{mm;j}$ and $F_2=
  z_{nj}^* z_{mj}  Q^{12}_{nm;j} + z_{nj} z_{mj}^*  Q^{12}_{mn;j}$.
 Eq.(50) is derived by first using eq.(18), followed by repeated partial integration. 
 Note $\int \delta_w \delta_{e,j} [L_E P_{N1}] {\rm d}e_j {\rm d}Z_j = N L_e P_w$. 
 Within approximation (20) for $Q$s, $A_1, A_2$ can  further be simplified 
which on substitution in eq.(50) give the diffusion of $P_w$ in a closed form:

 \begin{eqnarray}
  {\partial P_{w} \over\partial \Lambda_w}=
   {\partial^2  \over\partial {w}^2} 
\left[w(\Omega_1-4 w) \right] P_{w}
-  {\partial \over\partial w} \left[ \Omega_2 - b w \right] + N L_e P_w
 \end{eqnarray}
where $\Omega_1= |z_{mj}|_{typ}^2 + |z_{nj}|_{typ}^2 +
2 |z_{nj}|_{typ}^2 \delta_{nm} = 
2(1+\delta_{nm})   u_j^{typ} $, with 
$ u_j^{typ}={\rm exp}[\langle {\rm ln} u_j \rangle] $ as the typical  
local intensity of the $j^{th}$ eigenfunction, $\Omega_2=(\beta/2) 
[ |z_{mj}|^2_{typ} +  |z_{nj}|^2_{typ} +
(4/\beta)  |z_{nj}|^2_{typ} \delta_{nm}]
=(\beta+2\delta_{nm}) u_j^{typ} $,  $\Lambda_w = 2\chi \Lambda $ and $b=\beta N + 2$. 
The last term on the 
right of eq.(53) can be removed by an integration over energy $e$, leaving us with 
an evolution equation for energy-averaged $P_w$. Note the energy averaging of 
eq.(53) for case $n=m$ corresponds to eq.(46) for $P(I_2)$ (as $\sum_n w(n,n)=I_2$).

Exploiting the similarity of the form of energy-averaged eq.(53) to eq.(46), the 
behavior of $P_w(w)$ in different regimes can again be given by eqs.(48,49) after 
following replacements (everywhere in the equations): 
 $I_q \rightarrow w, \alpha_{in} \rightarrow \alpha_{wn}$ 
where $\alpha_{wn} \sim (-1)^n 2^{1-2n} \Omega_2^n ( w^{typ})^{-n} $ 
for $n \gtrsim 1$ and 
$\alpha_{w0} \approx b (1-{\rm e}^{-b\Lambda_w})^{-1}/4$.  
 Thus $P_w(w)$ decays exponentially 
for small $w$ ranges $(w \lesssim  w^{typ}/e )$: 
$P_w  \simeq  {\rm exp}\left[ - \alpha_{w0} w \right]$. 
It shows a power-law behavior for regimes 
$e^{-1} w^{typ} \gtrsim w \gtrsim  w^{typ} $   
$ P_w \simeq w^{-1} {\rm e}^{ -\alpha_{w1}{\rm ln} (\kappa w)}$,   
a log-normal decay for regimes  $w^{typ} \gtrsim w \gtrsim e w^{typ} $.    
Such a behavior was predicted by non-linear sigma model studies of 
quasi 1-d disordered wires too \cite{mir}.

Eq.(53) can be used to study the  behavior of various moments of the distribution 
of pair-correlation. For example, for average behavior of $w$, that is,   
$\langle w \rangle =\int w  P_w (w;\Lambda) \; {\rm d}w$, eq.(53) gives  
its $\Lambda$ evolution. The evolution equation turns out to be 
of the same form as eq.(36) (with $q=2$)  with following replacements:  
$\langle I_q \rangle \rightarrow \langle w \rangle$, $\alpha \rightarrow \Omega_2$ 
and $2 \Lambda \rightarrow \Lambda_w$ 
(note $\langle I_{q-1} \rangle=1$, $t \rightarrow b$ for $q=2$).  
%
It can be solved to show that 
$\langle w(\Lambda_w) \rangle = {\rm e}^{-b\Lambda_w} \left[ \langle w(0) \rangle + 
(\Omega_2/b)({\rm e}^{b\Lambda_w}-1)\right]$. 
A choice of $\langle w(0) \rangle =\delta_{ml}$ (corresponding to localized regime) gives 
$\langle w(\Lambda_w) \rangle =\delta_{ml}{\rm e}^{-b\Lambda_w}+
(\Omega_2/b)(1-{\rm e}^{-b\Lambda_w}) \approx (1-\kappa)/N $ 
which is analogous to the result obtained for disordered systems (by NLSM techniques); 
see \cite{mir}.

 \subsection {Correlation Between Eigenfunctions At Two Different Energies}

Critical point studies of many systems indicate the presence of 
multifractal structures among eigenfunctions.  
The multifractality suggests that the wavefunction is effectively located 
in a vanishingly small fraction of the system volume. However such extremely 
sparse wavefunctions can exhibit strong correlations if they belong to 
neighboring energy levels; the correlations therefore preserve the level-repulsion 
despite the sparsity of the wavefunction. Thus, for a complete analysis  
of level-statistics and associated physical properties, a knowledge of correlations 
among eigenfunction is very important. The correlations are also used 
in the analysis of many other physical properties e.g. for the 
measurement of the linear response of the system, or, to determine 
the fluctuations of matrix elements of some operator in a given basis. This information 
is useful in studies of the effect of a particular interaction on the statistical 
properties of the system e.g. effect of electron-electron interaction on a single 
particle disordered system.

 The correlations between components of two eigenfunctions at different 
 energies can be described as $\sigma(n,m,e_k,e_l)=|z_{nk} z_{ml}|^2$) 
(equivalently, in a continuous basis: $\sigma(r,r',e,e')=|\psi_e(r)\psi_{e'}(r')|^2)$. 
         The distribution $P_{\sigma}$ of the correlation $\sigma= 
 |z_{nk} z_{nl}|^2$ between $n^{\rm th}$ and $m^{\rm th}$ components  of the 
eigenfunctions $Z_k$ and  $Z_l$, respectively, is related to $P_{N2}$: 
${\bf P_{\sigma}(\sigma,e,\omega)= \sum_{k,l} \int \delta_{\sigma} \delta_{e,k} 
 \delta_{e+\omega,l} P_{N2}(Z_k,Z_l,e_k,e_l,Y) \; 
 \left(\prod_{j=k,l} {\rm d}\tau_j \right)}$
with $\delta_{\sigma}\equiv \delta(\sigma-|z_{nk} z_{ml}|^2)$, ${\rm d}\tau_j$ 
same as in eq.(20),  
 and, $\omega=|e_k-e_l|$ as the energy difference between two states. 
 Using eq.(25) and proceeding as in the case of $P_w$, the diffusion of 
$P_{\sigma}(\sigma)$  can be shown to be described by the equation

 \begin{eqnarray}
 {\partial P_{\sigma} \over\partial \Lambda_{\sigma}}=
   \left[ {\partial^2  \over\partial {\sigma}^2} 
 \left[\sigma (\tilde\Omega_1 -\sigma) \right] 
 -  {\partial \over\partial \sigma}\left[(\tilde\Omega_2-b \sigma) \right] 
+ L_e + L_{e+\omega} \right] P_{\sigma}
 \end{eqnarray}
where $\Lambda_{\sigma}=2 \chi \Lambda$, $\tilde\Omega_1=( {|z_{sl}|^2}_{typ} +
 {|z_{rk}|^2}_{typ} + 2  {|z_{rk}|^2}_{typ} \delta_{kl}\delta_{rs})$, 
$\tilde\Omega_2=[\beta( {|z_{rk}|^2}_{typ} + {|z_{sl}^2}_{typ})+ 
4 {|z_{rk}|^2}_{typ} \delta_{kl} \delta_{rs}]/2 $ 
and $b=\beta N+2$). Note eq.(53) is a special case of the above equation 
(as $P(w)\equiv P(\sigma)$ for $k=l$).

	The energy averaging of eq.(54) once again leads to an equation 
similar in form to eq.(46). Exploiting the analogy, we again get three different 
regimes for $P_{\sigma}(\sigma)$: (i) $\simeq  
{\rm exp}\left[ - \alpha_{\sigma 0} \sigma \right]$ (for    
$\sigma \lesssim  e^{-1} \sigma^{typ}$),    
(ii) $\simeq  {\sigma}^{-1} 
{\rm exp}{\left[ -\alpha_{\sigma 1} {\rm ln} (\sigma/\sigma^{typ})\right]}$ 
(for $e^{-1}\sigma^{typ} \gtrsim \sigma \gtrsim \sigma^{typ}$ and 
(iii) $ \simeq  {\sigma}^{-1} 
{\rm exp}\left[ -  \alpha_{\sigma 2} {\rm ln}^2 (\sigma/\sigma^{typ}) \right]$  
for ${ e \sigma^{typ} \sigma \gtrsim \sigma^{typ}}$ 
where $\alpha_{\sigma 0} \approx b (1-{\rm e}^{-b\Lambda_{\sigma}})^{-1}/4$ and  
$\alpha_{\sigma n} \sim (-1)^n 2^{1-2n} {\tilde\Omega_2}^n ( {\sigma}^{typ})^{-n} $ 
for $n \gtrsim 1$.

The $\Lambda$ dependence of average behavior of $\langle \sigma \rangle $ can now be 
derived by multiplying eq.(54) by $\sigma$ and then integrating over $\sigma$;
the equation again turn out to be same as the $q=2$ case of eq.(36) after 
following replacements:  
$\langle I_q \rangle \rightarrow \langle \sigma \rangle$, 
$\alpha \rightarrow {\tilde\Omega_2}$ and $2 \Lambda \rightarrow \Lambda_{\sigma}$.  
Solving the so-obtained evolution equation gives  
$\langle \sigma(\Lambda_{\sigma}) \rangle = {\rm e}^{-b\Lambda_{\sigma}} 
\left[\langle \sigma(0) \rangle  + (\tilde\Omega_2/b)({\rm e}^{b\Lambda_{\sigma}}-1)\right]$. 
 The  choice of a localized initial state (e.g. an insulator at 
$\Lambda_{\sigma}=0$) corresponds to $\langle \sigma(0) \rangle =0$ which gives 
$\langle \sigma(\Lambda) \rangle  \approx {\beta \over N} (1-{\rm e}^{-\beta \Lambda})$.

	In this paper, we have considered only two point correlations. 
The other correlations e.g. $\langle z^*_{rk} z_{rl} z_{sk} z^*_{sl} \rangle $ related to 
linear response of the system or, higher order ones e.g.
$\langle |z_{rk}|^4 |z_{sk}|^4 \rangle $ related to IPR fluctuations, can  similarly be 
determined using eq.(25).   
 
\subsection {Local Density of States $\rho(e,j)$}
 
	The local density of states or the spectral function, defined as   
  $\rho(e,j)=\sum_n |U_{jn})|^2 \delta(e-e_n)$, is an important measure 
of localization. This is because it counts the eigenstates $U_n$ having 
appreciable overlap with (or equivalently, located close to) the site $j$. 
Note this is distinct from the global density of states $\rho(e)$ which counts 
all the eigenstates at the energy $e$ irrespective of their location in space.
The measure $\rho(e,j)$ is  of special interest as it gives information 
about the decay of a specific unperturbed state into other states due to 
interaction. The width of the LDOS defines the effective "life-time" of the 
unperturbed basis state. Its distribution is an experimentally accessible 
quantity related to the position  and form of NMR line \cite{mir}.   
 
 The probability density $P_{\rho}(\rho)$ of $\rho(e,j)$ is related to $P_{1N}$: 
 ${\bf P_{\rho}(\rho)= \int \delta_{\rho} P_{1N}(E,Z,Y) \; {\rm d}E \; 
{\rm d^{\beta}}Z}$ 
 where $\delta_{\rho}=\delta(\rho-\sum_n |z_{jn}|^2 \delta(e-e_n))$ 
(note here $Z \equiv \{z_{jn}\}_{n=1,..N}$). 
 The diffusion of $P_{\rho}$ due to changing system parameters can again be 
 studied with the help of eq.(14) for $P_{1N}$. 
 \begin{eqnarray}
 {\partial P_{\rho} \over\partial Y}= L_E P_{\rho} + B+B^*   
 \end{eqnarray}
with $B= B^* =\int \delta_{\rho}\; L_z P_{1N} \;{\rm d}^{\beta}Z {\rm d}E 
=  {\partial^2 F_1 \over\partial \rho^2} + 
{\beta\over 2} {\partial F_2 \over\partial \rho}$. 
Here the $2^{nd}$ form of $B$ is obtained from the $1^{st}$ by 
a substitution of three terms of $L_z$ (eq.(15)) in the integral, and, a 
subsequent partial integration which gives       
\begin{eqnarray}
F_1 &=& \int \delta_{\rho} 
\left[\sum_{m,n;m\not=n}  {|z_{jn}|^2 |z_{jm}|^2\over (e_n-e_m)^2} 
\delta(e-e_n) \left(1- \delta(e-e_m)\right) \right]
 P_{1N} \;{\rm d^{\beta}}Z {\rm d}E \\
F_2 &=& \int \delta_{\rho} \left[  \sum_{m,n;m\not=n} 
{|z_{jm}|^2-|z_{jn}|^2 \over (e_n-e_m)^2} \delta(e-e_n) \right]
 P_{1N} \;{\rm d^{\beta}}Z {\rm d}E 
 \end{eqnarray}

As in the case of the integrals $Q$ and $G$ (see Appendix A),  
the dominant contribution to the integrals $F_1$ and $F_2$ comes from 
the regions where the exponent term in $P_{1N}$, that is 
$f= (\mu/2) \sum_{m<n} |e_n-e_m|^2 |U_{jn}|^2 |U_{jm}|^2 < 1$. 
This can occur under two conditions: 

(1) $\mu \langle \sigma \rangle  < 1$: Here $\sigma=|U_{jn}|^2 |U_{jm}|^2$ describes the 
correlation between two different eigenfunction components in the same basis state. 
Under the condition, a neighborhood of the order of mean level spacing can contribute to 
the integral over e-variables i.e $|e_n-e_m|^2 \sim D^2$, 

(2) $\mu \langle \sigma \rangle  \ge 1$: 
In this case, $f \gg 1 $ only for those regions where 
 $|e_n-e_m|^2 \sim D^2/(\mu \langle \sigma \rangle )$.  
Note however in both cases, almost entire eigenfunction space can contribute 
to the integral. 

Thus $F_1$ and $F_2$ can be approximated as
$F_1 \approx {\chi D^{-2}} \left[ \rho(1-  \rho)P_{\rho}  \right]$ 
and $F_2 \approx N \chi D^{-2} \left[\rho-\langle \rho \rangle \right]P_{\rho}$ where 
$\chi=1$ if $\mu \langle \sigma \rangle  \lesssim 1$ and 
$\chi \sim \mu \langle \sigma \rangle$ for $\mu \langle \sigma \rangle  \gtrsim 1$. 
The approximate forms of $F_1$ and $F_2$ can now be used to rewrite $B$ as a function 
of $P_{\rho}$ which on substitution in eq.(55)  gives the diffusion equation 
for $P_{\rho}$: 
     
 \begin{eqnarray}
 {\partial P_{\rho} \over\partial \Lambda_{\rho}}= L_e P_{\rho} + 
  {\partial^2  \over\partial \rho^2} \left[ \rho(1- \rho)  \right] P_{\rho} 
 + {\beta N\over 2} {\partial  \over\partial \rho} 
\left[\rho-\langle \rho \rangle \right]P_{\rho}
 \end{eqnarray}
where $\Lambda_{\rho}=\chi \Lambda$. Note that the above equation is analogous in 
form as  the eqs.(46, 53, 54) of $P_{I_q}$, $P_w$ and $P_{\sigma}$, respectively. 
This similarity is reflected in 
both short as well as  long-range behavior of $P_{\rho}$:
(i)   $\simeq  {\rm exp}\left[ - \alpha_{\rho 0} \rho \right] $  
(for $\rho \lesssim  \rho^{typ}$), 
(ii) $ \simeq  {\rho}^{-1} {\rm exp}{ -\alpha_{\rho 1} {\rm ln} (\kappa \rho)}$
(for $\kappa_1^{-1} \gtrsim \rho \gtrsim   \rho^{typ}$), and, 
(iii) $ \simeq  {\rho}^{-1} {\rm exp}\left[ -  \alpha_3{\rm ln}^2 (\kappa \rho) \right]$   
 for (${\sigma \gtrsim \kappa^{-1}}$) 
where $\alpha_{\rho 0} \approx b (1-{\rm e}^{-b\Lambda_{\rho}})^{-1}/4$ and  
$\alpha_{\rho n} \sim (-1)^n 2^{1-2n} ( {\rho}^{typ})^{-n} $.

  . 
 
\section{The Parameter $\Lambda$}
 
         The set of Eqs.(14,18,25)   provides a common 
 mathematical formulation for the eigenfunction-statistics of various 
 complex systems modeled by eqs.(1,2);  here the information about the system enters 
 only through $Y$. As shown explicitly in \cite{ps1}, the same $Y$ also enters in the 
 common mathematical formulation of the eigenvalue-statistics of ensemble (1) 
(and (2), see \cite{ps3});  this is implied by eqs.(14,18,25) too. However, 
as discussed in \cite{ps1, ps3}, the 
 evolution of the $n^{\rm th}$ order eigenvalue correlations ($n>1$) as a function 
 of $Y$, is abrupt in large $N$-limit; a smooth crossover can only be seen in terms 
 of a rescaled parameter $\Lambda_e$ where
 
 \begin{eqnarray}
 \Lambda_e (e,Y)= \Lambda={Y-Y_0\over D_{\zeta}^2}
 \end{eqnarray} 
 with $D_{\zeta}(e,Y)=D (\zeta/L)^d$ as the local mean level spacing, $D(e,Y)$ as the 
 mean level spacing of the full spectrum and $\zeta$ as the correlation/localization 
 length for a $d$-dimensional system of length $L$ ($N=L^d$), at an energy $e$ and 
 parameter $Y$ (with $Y_0$ as its initial value). Thus $\Lambda_e$ for various systems e.g 
disordered systems, mixed systems, systems with chiral or particle-hole symmetry etc. 
can be calculated by a prior knowledge of system parameters; (e.g. see \cite{ps2} for  
the calculation  for Anderson and Brownian ensembles). As $\Lambda_e$ increases from 
zero to infinity, the level-statistics changes from its initial state (with $Y=Y_0$) 
to that of Wigner-Dyson ensemble. For example, let the initial state corresponds to 
insulator limit of disordered systems or integrable limit of mixed systems; both limits 
show  Poisson level-statistics \cite{du, mir, be2, be3, gu}. A variation of system 
parameters changes $\Lambda$ 
from zero, causing diffusion of levels towards Wigner-Dyson steady state. According to 
$\Lambda$ formulation, the  level-statistics, for the system parameters resulting in 
finite $\Lambda$, is then an intermediate point of Poisson $\rightarrow$ Wigner-Dyson 
transition. The prediction is in agreement with previous works  on the two systems 
\cite{mir, iz, ps2, be2, be3}; (note Wigner-Dyson statistics corresponds 
to metal and chaotic limits of disordered and mixed systems, respectively 
\cite{du, bohi}).

	For later reference, it is worth reviewing the role of $\Lambda_e$, that is, 
$\Lambda$ in locating the critical point of level statistics. As both $|Y-Y_0|$ as 
well as the local mean level density are functions of $N$, the latter can affect $\Lambda$ 
significantly. As a consequence, the size $N$ plays a crucial role in determining the 
level statistics in the critical regime. For finite systems, the eigenvalue statistics 
smoothly approaches  one of the two end points, namely, $\Lambda\rightarrow 0$ or 
 $\Lambda \rightarrow \infty$, with increasing system size. 
 The variation of $\Lambda$ in infinite systems, however, may lead to an abrupt 
transition of the statistics,  with its critical point given by the condition 
$\Lambda=\it {size \; independent}$ (see ref.\cite{mj} for the definition of 
a critical distribution).  The finite, non-zero $\Lambda$ strength, say 
$\Lambda_{critical}$, at the critical point, 
results in an eigenvalue-statistics different from the two end points. 
Note, however, that the existence of a critical point or its absence depends on the 
relative size-dependencies of $|Y-Y_0|$ and the local mean level spacing. If the 
size-dependence of $D_{\zeta}^2$ remains different  from that of ${|Y-Y_0|}$ under all 
complexity conditions, $\Lambda$ will never achieve a finite non-zero value in infinite 
size limit. As a consequence, such a system will not show a critical behavior of 
eigenvalue-statistics. For example, as discussed in \cite{ps2} for a $d$-dimensional 
Anderson Hamiltonian of linear size $L$, $\Lambda$ turns out to be size-independent 
only for $d>2$. The $\Lambda$-formulation, therefore, indicates the lack of 
metal-insulator transition for dimensions $d \le 2$ which is in agreement with 
several studies of previous years.

 The connection of the eigenvalue fluctuations to those of eigenfunctions 
 suggested $\Lambda$ as the  evolution parameter for the eigenfunctions correlations 
 (of order $n>1$) too. As shown in section III, the evolution parameters 
$\Lambda_{measure}$  of various eigenfunction fluctuation measures are indeed functions   
of $\Lambda$: $\Lambda_{measure}=f(\Lambda)$. Here $f(\Lambda) \propto \Lambda$ on short 
length scales and $f(\Lambda) \sim \Lambda {\rm e}^{-\alpha \Lambda}$ in the tail regime. 


 The parameter $\Lambda$, being a function of the distribution parameters of the 
 matrix elements, is sensitive to changes in the system parameters; this is 
 due to latter's influence on the uncertainties associated with system-interactions. 
 Some examples of such system parameters are disorder, dimensionality, boundary 
 and topological conditions, system size etc. For example, the presence of disorder 
 randomizes the interactions,  with degree of disorder 
 affecting the distribution parameters $h,b$ and consequently $\Lambda$.  
The dependence of $\Lambda$ on the dimensionality and  boundary 
conditions originates from their influence on the basis 
connectivity i.e degree of sparsity of the matrix which is reflected in the 
distribution parameters $h,b$. For example, for nearest neighbor hopping and hard wall 
boundary conditions in $d$-dimensions, the matrix element $H_{jk} \not=0$ 
only if $j=|k\pm L^{d-1}|$ (with $L$ as linear size). The variance $h_{jk}$ of the 
distribution $\rho(H_{jk})$ is 
therefore finite only for $j=k$ or $|j=k\pm L^{d-1}|$ and is zero for all other $j,k$.   
 The information about dimensionality in $\Lambda$ also enters through    
the local mean level spacing which depends on the correlation volume $\zeta^d$. 
 (See also \cite{ps2} where the dependence of $\Lambda$ on 
 system parameters is explained  by considering an example of Anderson Hamiltonian). 
 
	The system size $N$ is another important parameter which affects the evolution 
of the measures significantly. As shown in section IV, it appears independently as 
well as through $\Lambda_{measure}$ in the evolution equations which suggests a two 
parametric dependence, namely, $\Lambda_{measure}$ and $N$ (separately) of these measures. 
As a consequence, even at the critical point of level statistics, 
the eigenfunction statistics remains sensitive to size $N$. 
This in turn results in  a multifractal behavior of the eigenfunctions at the critical 
point of any complex system, modeled by eqs.(1,2). The scaling exponents at the critical 
point, referred as critical exponents or multifractal dimensions, depend on system 
parameters. In finite size systems, changing system parameters can change $\Lambda$ (and therefore 
$\Lambda_{measure}$) continuously between $0$ and $\infty$ which may lead to intermediate 
stages of varying degree of multifractality. However, the physically interesting cases 
usually correspond to infinite sizes where $\Lambda$ takes only   
three possible values, namely, $\Lambda=0, \infty, \Lambda_{critical}$; for these cases 
therefore only one multifractal stage, that is at the critical point corresponding 
to $\Lambda_{critical}$, can exist. As $\Lambda_{critical}$ is 
sensitive to system-specifics, the critical (multifractal) exponents can vary from 
system to system. Note, as already mentioned above, the occurrence of 
critical point and, therefore, a multifractal behavior of eigenstates is not a necessary 
feature of all infinite size complex systems. 


 The $\Lambda$-governed diffusion equations, derived in section IV, are valid 
 for arbitrary initial conditions at $\Lambda=0$ (which implies $\Lambda_x=0$) and 
their solutions  $P_x(X,\Lambda_x|X_0,0)$ describe the probability of the measure, 
say $X$, at  $\Lambda_x$ for a given initial state of $X=X_0$. Thus $P$ is subjected to 
an initial constraint ${\rm lim \Lambda \rightarrow 0} P(X,\Lambda_x|X_0,0) 
=\delta(X-X_0)$. By integration of the solution over the distribution of initial 
values  $P_0(X_0,0)$, one can recover  $P(X,\Lambda_x)$, that is, the distribution of 
measure $X$ for a system with  complexity parameter strength $\Lambda$:
 \begin{eqnarray}
 P(X,\Lambda_x)=\int P(X,\Lambda_x|X_0,0) P_0(X_0,0) {\rm d}X_0. 
 \end{eqnarray}


 The eq.(60) implies that the statistics evolved in "time" $\Lambda_x$ 
 is sensitive to the collective behavior of system parameters contributing to 
 $\Lambda_x$  and the initial distribution only. The latter can 
 always be chosen same for the systems operating in the  matrix spaces of 
 similar type e.g. same symmetry conditions; (the initial values of their 
 $Y$ parameters need not be equal). Thus if both A,B operate in the matrix spaces 
 of same type, their behavior at 
 the system parameter strengths which lead to $\Lambda_{x,A}=\Lambda_{x,B}=t$ would also 
 be same (although they may show different behavior between 
 $0<\Lambda_{x,A}, \Lambda_{x,B} < t$). This implies a great deal of universality among 
 systems of widely different origins of complexity. For example, consider the cases of a 
 three dimensional disordered system, say A, and 
 a clean, closed quantum dot, say B. In the first case, $\Lambda_{x,A}=\Lambda_{x,disorder}$ is a 
function of  disorder, hopping strength, dimensionality, boundary condition etc.  
  In the case of a dot, $\Lambda_{x,B}=\Lambda_{x,dot}$ is a function of shape as well as 
size. It is well-known that, in strong disorder limit 
 and  for circular shape, respectively, both  systems show localized wavefunction 
 dynamics and same statistical behavior of the eigenfunctions and eigenvalues of 
 the Hamiltonians. Reducing the degree of disorder or change of shape of the dot 
from circle to  stadium type results in a transition from localized to delocalized 
dynamics  of the wavefunctions. The statistics in the intermediate stages during the 
transition for each case  is governed by the respective $\Lambda_x$ strengths. 
 If, however, the $\Lambda_{x,dot}=\Lambda_{x,disorder}$ at some shape parameter 
 and disorder strength, respectively,  our analysis predicts 
 a same statistical behavior for both systems. The 
 implication can also be extended to classical systems e.g. stock market 
 fluctuations which are analyzed by statistical studies of the correlation 
 matrix of stocks \cite{pl}. (Note, the correlation matrices of classical systems 
are, in general, non-Hermitian; however, as shown in \cite{ps0}, the $\Lambda$ 
formulation remains valid for non-Hermitian version of eq.(1)).    
Here a very weak interaction among certain 
 stocks due to various socio-economic conditions results in a localized dynamics 
 of the eigenfunctions. The changing conditions 
 may lead to a more homogenized interaction of some of the stocks, thus 
 introducing a transition from localized to delocalized wave dynamics. 
 In this case, $\Lambda$  is a function of the socio-economic 
 parameters (SEP). However if $\Lambda_{x,stock}=\Lambda_{x,dot}$ 
 for some combinations of SEP and dot parameters, respectively, the spectral and strength 
fluctuations in correlation matrix of the stock market and the Hamiltonian of 
quantum dot will show same behavior. Note the analogy of statistical behavior 
of the eigenvalues and eigenfunctions among the three systems, mentioned above, has 
already been numerically verified in delocalized waves limit 
$\Lambda \rightarrow \infty$ \cite{mir, alh, pl}.  
.  
 
 The above universality makes $\Lambda$ formulation useful as it can be 
 exploited to obtain the statistics of a complex system if the same information 
 is available about another system (under same symmetry conditions) by another 
 method. For example, for Anderson type  disordered Hamiltonian, the distributions of 
many measures  are known by non-linear sigma model techniques. The 
formulations can then be  used for the complex systems e.g stock markets undergoing a 
transition from localized  $\rightarrow $ delocalized wave dynamics;
one just needs to replace $\Lambda({\rm Anderson})$ by that of the system. 

  The formulation can also be used to search for the system 
conditions leading to a critical state or multifractal wavefunctions of 
various complex systems. For example, the suggested modelling of mixed systems by 
eq.(1) would imply the possible existence of a critical point of 
level-statistics in the systems and multifractal eigenstates. 
The intuition suggests that the occurrence of such a point  may correspond to breaking of 
the last KAM curve, thus allowing classical diffusion or delocalization of the 
dynamics above the critical point and localization below it; 
however it needs to be further explored.   
The critical $\Lambda$ can then given by the critical value of system parameter 
leading to last KAM curve breaking.

 \section{Numerical Analysis}
 
 For numerical analysis, we choose three different ensembles (for both cases $\beta=1,2$); 
 the choice is dictated by the reason (i) the ensembles are prototype models 
 of many physical systems related to different areas \cite{mir, me, gu, gg}, 
  (ii) a comparative study of the eigenvalue fluctuations of these systems has already 
been carried out, with their $\Lambda$  parameters and other results given in \cite{ps1}:

 (i) Critical Anderson ensemble (AE): 

 (a) {\it Time-Reversal case} ${\rm AE}_t$: 
 we analyze cubic ($d=3$) Anderson lattice of linear size $L$ ($N=L^d$) with a Gaussian site 
 disorder (of variance $W^2/12$, $W=4.05$ and mean zero), same for each site, 
 an isotropic random hopping between nearest neighbors  
 with hard wall boundary conditions \cite{mir,ps2}. 
 The ensemble density in this case can be described by eq.(1) with  
$h_{kk} = W^2/12, h_{kl} = f(kl)/12, b_{kl}=0$ 
where $f(kl)=1$ for $\{k,l\}$ pairs representing hopping,  $f(kl)\rightarrow 0$ 
for all $\{k,l\}$ values corresponding to disconnected sites. 
 A substitution of above values in eq.(8) gives $Y$ which subsequently gives 
$\Lambda$ by eq.(59): 
 (see eq.(19) of  \cite{ps1}), 
 \begin{eqnarray}
 \Lambda_a(E,Y) &=& |\alpha-\alpha_0| F^2 \zeta^{2d} L^{-d} \gamma^{-1}
 \end{eqnarray}  
 with $\alpha-\alpha_0=1.36$ and  $F(E)=0.26 e^{-E^2/5}$ 
 (see section V of \cite{ps1}). (Note, for later reference, 
$F(E)$ is the mean level density: $F(E)=(N D)^{-1}$).

(b) {\it Broken Time-Reversal case} ${\rm AE}_{nt}$: 
 we analyze cubic ($d=3$) Anderson lattice of linear size $L$ ($N=L^d$) with a 
 Gaussian site disorder (of variance $W^2/12$, $W=21.3$ and mean zero), same for 
 each site, an isotropic non-random hopping $t=1$ between nearest neighbors  
 with periodic boundary conditions \cite{mir,ps2}. 
The time-reversal symmetry is broken by 
applying an Aharnov Bohm flux $\phi$  which gives 
rise to a nearest neighbor hopping $H_{kl}={\rm exp}(i \phi)$ 
for all $k,l$ values related to the nearest-neighbor pairs \cite{ter}.   
The flux $\phi$ is chosen to be non-random in nature, that is, 
$\langle {\rm cos}^2(\phi) \rangle =W_1=0$, $<{\rm sin}^2(\phi)>=W_2=0$ and  
$\langle {\rm cos}(\phi) \rangle =t_1=1$, $<{\rm sin}(\phi)>=t_2=1$.   
 The ensemble density in this case can again be described  by eq.(1) with  
$ h_{kk} = W^2/12,  b_{kk}=0, h_{kl;s} = W_s=0,  b_{kl;s}= t_s f(kl;s)$ 
where $f(kl;s)=1$ for $\{k,l\}$ pairs representing hopping,  $f(kl;s)\rightarrow 0$ 
for all $\{k,l\}$ values corresponding to disconnected sites. 
The $\Lambda$ for this case is still given by  eq.(61) (except for a 
factor $\beta^{-1}$); however now $\alpha-\alpha_0=5.43$, $F(E)=0.016 e^{-E^2/400}$ 
 (see section V of \cite{ps1}).

 (ii) Critical Power Law Random Banded Ensemble (critical PRBM or PE): 
As mentioned in section II, PRBM ensemble was introduced as a possible model 
for the  level statistics of Anderson Hamiltonian  \cite{mir1}. It 
is defined as the ensemble of random Hermitian matrices with matrix elements 
$H_{ij}$ as independently distributed Gaussian variables with zero 
mean i.e $<H_{ij}>=0$ and a power-law decay of the variances 
away from the diagonal \cite{mir1, mir, fyd}: $<|H_{ij;s}|^2>=a(|i-j|)$  
with function $a(r)$ decaying as $r$ increases.

The PRBM ensemble with specific choice 
$<|H_{ij;s}|^2>= G_{ij}^{-1} \left[ 1+ (|i-j|/b)^2 \right]^{-1} $, 
$G_{ij}=\beta (2-\delta_{ij})$ and 
$G_{ij}=1/2$ (referred as critical PRBM or PE in this paper)  
leads to a critical behavior of eigenfunction and eigenvalue statistics 
at arbitrary values of the parameter $b$ 
and is believed to show all the key features of the Anderson critical 
point, including multifractality of eigenfunctions and the 
fractional spectral compressibility \cite{mir1, mir}.  
The ensemble density in this case corresponds to eq.(1)    
with  $b_{kl}=0$, and, $h_{kl;s}= G_{kl}^{-1} \left[ 1+ (|k-l|/p)^2 \right]^{-1} $. 
 The corresponding $\Lambda$ can be shown to be given by  
 (see section VI of \cite{ps1}),
 \begin{eqnarray}
 \Lambda_{p}(p,E) = \alpha_p^{-1} f(p) F^2(E) \zeta^2 N^{-1}. 
 \end{eqnarray}
 where $\alpha_p=2N (N+2-\beta)$, $f(p)=\sum_{r=1}^N (N-r) {\rm ln}|1+(p/r)^2|$.  


 (iii) Critical Brownian Ensemble (BE): 
A Brownian ensemble can be described as a non-stationary 
state of the matrix elements undergoing a cross-over due to a random perturbation 
of a stationary ensemble by another one \cite{dy,me,ap,ps1}. For example, 
in the case of Hermitian operators, a Brownian ensemble $H$ can be given 
as $H=\sqrt{f} (H_0+\lambda V)$ (with $f=(1-\lambda^2)^{-1}$); here $V$ is a 
random perturbation of strength $\lambda$, taken from a stationary ensemble 
\cite{zirn} e.g. Wigner-Dyson ensemble, and applied to an initial stationary 
state $H_0$ (see also \cite{ap}). Here we consider a specific class of BEs, 
namely, those appearing during a transition from Poisson $\rightarrow$ 
Wigner-Dyson ensemble, caused by a perturbation of the former by 
the latter (that is, taking $H_0$, $V$ as Poisson and Wigner-Dyson ensemble 
respectively). As, in above two cases,  this transition also results in a change 
of localized eigenstates to delocalized ones. The BEs related to the Poisson $\rightarrow$ 
Wigner-Dyson transition can be described by a $N\times N$ ensemble $H$ 
represented by eq.(1) with mean $\langle H_{kl} \rangle =b_{kl}=0$, the variance 
$\langle H_{kk;s}^2 \rangle = h_{kk;s}=(2\gamma)^{-1}$ and 
 $\langle H_{kl;s}^2 \rangle = h_{kl;s}=[4\gamma (1+ \mu)]^{-1}$ for $k\not=l$. 
%
%
 with $(1+\mu)=(\lambda^2 f)^{-1}$; here $H=H_0$ for $\lambda \rightarrow 0$ or 
$\mu \rightarrow \infty$. As mentioned in section II, 
the ensemble density in this case has the same form as for Rosenzweig-Porter (RP) 
ensemble \cite{rp}; it can also describe  an ensemble of Anderson 
Hamiltonians with very long range, isotropic, random hopping.
Further, as discussed in \cite{ps1}, the special case $\mu=cN^2$ 
corresponds to the critical BEs; their mean level density is given as 
$F(E)=(\pi)^{-1/2} e^{-E^2}$ and   
 \begin{eqnarray}
 \Lambda_b(E) = (1/4 c\pi\gamma){\rm e}^{-E^2}. 
 \end{eqnarray}

 
         Our aim is to show that the behavior of an eigenfunction fluctuation measure of 
 AE, BE and PE is analogous at system parameters which lead to a same $\Lambda_{measure}$ 
 value for all the three cases. Using the latter as a condition, we can  obtain the desired 
system parameters in each case (that is, $p$ for PE and $c$ for BE for  a given AE). 
 As $\Lambda$ for the three cases is energy dependent, the fluctuation measures should be 
compared at precisely a given value of energy.  For numerical analysis, however, one needs 
to consider averages over an energy range  $\Delta E$ which should be sufficiently large in 
order to improve the statistics. On the other hand, choice of a very large $\Delta E$ 
will lead to mixing of different statistics (in a range $\Delta \Lambda \propto \delta E$). 
 As a consequence, one needs to consider an optimized range of $\Delta E$. In our 
simulations, we analyze large ensembles of about 1400 matrices of size $N=2197$. 
We choose $\Delta E$ to be about 10${\% }$ of the bandwidth, at the band 
center $E=0$ which gives approximately $3\times 10^5$ levels for each case. 
As the chosen $\Delta E$ corresponds to a 1 $\%$ variation of the density of states, 
it avoids mixing of different statistics. 

	As discussed in previous section, the eigenfunction fluctuations 
are influenced by both $\Lambda$ as well as system size $N$. To compare  
$\Lambda_{measure}$ dependence of a fluctuation measure (of eigenfunctions), therefore, 
same system size should be taken for all systems under consideration. As examples, 
here we consider distributions of three measures, namely, local eigenfunction intensity 
$P_u(u)$, inverse participation ratio $P_I(I_2)$ and pair correlation function $P_w(w)$ 
for the three systems under time-reversal symmetry ($\beta=1$) i.e . 
${\rm AE}_t$, ${\rm BE}_t$, ${\rm PE}_t$. 
As, for $P_u$, 
\begin{eqnarray}
\Lambda_u = {\mu (Y-Y_0) \zeta^{2d} \over N^2 D^2} 
\approx {\left(F\over I_2^{typ}\right)}^2,
\end{eqnarray}
(with $\zeta^d \approx (I_2^{typ})^{-1}$), the BE and PE analogs for the intensity 
distribution of ${\rm AE}_t$ can be obtained by the condition 
$I_{2,a}^{typ}/F_a=I_{2,b}^{typ}/F_b=I_{2,p}^{typ}/F_p$. This requires a prior 
information about $I_2^{typ}$. Our numerical study for various sizes of the three systems   
shows that, for each case, $I_2^{typ} \approx {\tilde I} N^{-D_2}$ with ${\tilde I}$ 
and $D_2$ system dependent. 
The numerical information about $ I_2^{typ}$ and $F$  can now be used to obtain the 
parameters $p$ and $c$ for PE and BE analogs of ${\rm AE}_t$ for $P_u$ case 
(i.e PE and BE with the ratio $I_2^{typ} /F$ same as 
for AE); we find $p=0.4$, $c=0.02$. The figure 1 shows the distribution $P_u (u')$, 
 $u'=[{\rm ln}u-\langle {\rm ln}u \rangle]/ \langle {\rm ln}^2 u \rangle$, 
for the ${\rm BE}_t$ case 
($c=0.02$) and ${\rm PE}_t$ case ($p=0.4$) along with ${\rm AE}_t$ case;     
the close agreement among the three cases confirms our theoretical prediction. 
This is also confirmed by the comparison of $P_I({\rm ln} I_2)$ and 
$P_w({\rm ln} w)$ for the three systems, shown in figure 3 and figure 5, 
respectively. Here again the parameters $p$ and $c$ 
for PE and BE analogs for both measures are obtained by the relation  
$\Lambda_{I,a}=\Lambda_{I,b}=\Lambda_{I,p}$ (similarly for $w$). 

 The above numerical analysis is repeated also for the case of AE in a magnetic field and 
its BE and PE analog; the results for the three measures, shown in figures 2,4 and 
6, further support our claim: {\it the  eigenfunction fluctuations of different complex 
systems show same behavior if their complexity parameters and sizes are equal}. 
It is worth reminding that {\it behavior of the eigenvalue fluctuations is governed only 
by the related complexity parameter (that is, no independent influence of size)}. 
The details of analytical and numerical evidence about the eigenvalue statistics 
are already published in \cite{ps2,ps1,ps3}. However  
for the sake of completeness and to convince the 
reader, we include here the numerical analysis of an eigenvalue fluctuation measure, 
namely, nearest neighbor spacing distribution $P(S)$ for the three system 
(for both $\beta=1,2 $ cases) at parametric values leading 
to $\Lambda_{e,a}=\Lambda_{e,b}=\Lambda_{e,c}$ (where $\Lambda_e =\Lambda$); 
the plots shown in figures 7,8 reconfirm the claim about eigenvalue statistics. 


 

 \section{Conclusion}

	In the end, we summarize our main results. Our analysis of the eigenfunction 
correlations of complex systems indicates a two parameter dependence, namely,  
complexity parameter $\Lambda$ and system size $N$, 
of the distributions of eigenfunction components. The independent appearance of size 
parameter (besides through $\Lambda$) seems to suggest a lack of finite size scaling 
in eigenfunction distributions and an absence of their critical limit.
This is in contrast with the behavior of eigenvalue distribution which shows a 
single parametric scaling  as well as a critical limit if the condition  
${\rm lim} N\rightarrow\infty \; \Lambda={\rm finite}$ is satisfied by the system.
Note the above implies the size-dependence of eigenfunction correlations at the critical 
point of level-statistics too.   
   



      We have also studied the distribution of a few important measures of eigenfunction 
correlations e.g. local density of states, pair correlation function etc. 
We find that the form of complexity parameter governing an eigenfunction 
fluctuation measure is sensitive to its nature  (e.g. $\Lambda_u$ for local 
intensity distribution, $\Lambda_I$ for inverse participation ratio distribution etc). 
This is again different from the 
eigenvalue fluctuations (except for level density) which are all governed by the same 
complexity parameter, namely, $\Lambda_e=\Lambda$.  Our analysis 
indicates a log-normal behavior of the asymptotic tails of the distributions at finite 
$\Lambda$-strength. In context of disordered systems, a similar behavior was predicted 
by other studies using different techniques e.g. Berezinski and Abrikosov-Ryzkhin 
technique (for one dimension) and by 
non-linear sigma model (for higher dimensions) \cite{mir}. However the complexity parameter 
formulation suggests the existence of such a tail-behavior and multifractal 
eigenfunctions for almost any complex system,  
 irrespective of the origin of complexity, if the parameter $N\Lambda_{measure}$ is finite. 
A recent numerical study of the eigenfunction of the correlation matrix of 
stock prices confirms the suggestion in case of stock market \cite{pl,kwa}. 
As finite $\Lambda$ corresponds to the critical point condition in 
infinite size systems, a log-normal tail-behavior seems to be associated 
with the existence of a critical point (and vice-verse). The above study can thus be 
used to search and predict the critical stages of other complex systems e.g. 
stock market, brain etc.

	In this paper, we have considered the cases  modeled by  generalized Gaussian 
ensembles with  uncorrelated matrix elements as well as a wide range of non-Gaussian 
ensembles with correlated matrix elements. The latter are  suitable models, 
 for example, for disordered systems with varying degree of particle-particle 
 interactions. In context of disordered systems, therefore, we expect a same 
 statistical behavior of a measure for both the cases, namely, with or without particle 
interactions,  if the strengths of their parameters $\Lambda_{measure}$ are equal. 
 This  suggests the sensitivity of the statistical behavior of a disordered system to 
degree of its complexity only (measured by complexity parameter), irrespective of the origin.   
The statement is  expected to be valid for correlated and uncorrelated cases of other 
complex systems too. This in turn would indicate the  existence of an infinite family of  
 universality classes, parametrized by $\Lambda$, of statistical behavior among 
 complex systems.

\begin{appendix}

\section{Calculation of Integrals $Q_{mn;k}^{rs}$ and $G_r$}

The integral $Q_{mn;k}^{rs} $ defined by eq.(20) can be rewritten in terms of 
$\rho(H)$ as 

\begin{eqnarray}
Q_{mn;k}^{rs} &=& \sum_{j; j \not=k} \int {(U_{nj} U^*_{mj})^r \over 
(\lambda_k-\lambda_j)^{s}} \;{\tilde f_k} \; \rho (H,Y)\;  {\rm d}H 
\end{eqnarray}

To express $Q$ in terms of $P_{N1}$, it is necessary to write $\rho(H)$   
in eigenvalue-eigenvector space i.e. $\{\lambda,U\}$ space. The steps can briefly 
be given as follows. The solution of eq.(6) for arbitrary initial condition, 
say $H_0$ at $Y=Y_0$ can be  given as  
 $\rho(H,Y|H_0,Y_0) \propto {\rm exp}[-(\alpha/2) {\rm Tr}(H- \eta H_0)^2]$ with 
 $\alpha=\gamma (1-\eta^2)^{-1}$ and $\eta={\rm e}^{-\gamma Y}$.  
Without loss of  generality, the basis space for $H$ can be chosen as the eigenvector 
space of $H_0$; The initial ensemble $H_0$ in this basis consists of diagonal 
matrices. For simplification, consider the initial distribution given by  
$\rho(H_0) \propto {\rm e}^{-\sum_{j=1}^N H_{0;jj}^2}$.  
 Using eigenvalue equation $U H=\Lambda U$, $\rho(H,Y|H_0,Y_0)$ can be transformed from 
 matrix space to eigenvalue-eigenvector space $\{\lambda,U\}$ which followed 
by an integration over ensemble $H_0$ gives 

\begin{eqnarray}
 \rho(H,Y)  \propto \prod_{k,l; k<l}^N |\lambda_k-\lambda_l|^{\beta} 
 {\rm exp}\left[-(1/2) \sum_{j=1}^N \lambda_j^2 - 
 (\mu/2) \sum_{k<l} |\lambda_k-\lambda_l|^2 U_{jk}^2 U_{jl}^2 \right]
\end{eqnarray}
 where $\mu=({\rm e}^{2\eta(Y-Y_0)}-1)^{-1}$.

 A substitution of eq.(A2) for $\rho$ in eq.(A1) and using 
${\rm d}H =\prod_j \prod_{k<l} |\lambda_k-\lambda_l|^{\beta} {\rm d}\lambda_j {\rm d}U_j$, 
gives $Q$ as a function of $\{U,\Lambda\}$ variables. 
  As eq.(A2) indicates, the behavior of $Q_{mn;k}^{rs}$ is significantly 
influenced by the term $R \equiv \mu |\lambda_k-\lambda_j|^2 
\sum_{n=1}^N |U_{nk}|^2 |U_{nj}|^2$  present in the exponent of $\rho$. 
Consequently, for a given $Y$, the dominant contribution to the integrals over 
the variables $U_j$ and $\lambda_j$ in eq.(A1) comes from those regions which lead to 
$R \rightarrow 0$. Also note that the eigenvalue-eigenfunction correlations appear 
in $\rho$ only through $R$. The limit $R \rightarrow 0$ therefore  
allows a mutually independent integration over  $\lambda_j$ and $U_{nj}$ variables. 
As the typical local intensity $|U_{nk}|^2_{typical} \sim \zeta_k^{-d}$ with 
$\zeta_k$ as the localization length of the eigenfunction $U_k$ ($d$ as system-dimension),
this implies $R \sim \mu \zeta_k^{-d}|\lambda_k-\lambda_j|^2 $. Consequently,  
the regions of variable $\lambda_j$ and $U_{nj}$ which contribute to integral 
depend on mutual competition between $\mu$ and $\zeta_k^d$:   

(i) for $ \mu <\zeta_k^d$, almost entire region of $U_{nj}$ can 
    contribute to  integral (due to $0< |U_{nj}|^2 < 1$). However only a small 
    neighborhood of the order of local mean level spacing,  i.e 
    $|\lambda_k-\lambda_j| \simeq D_k$ around $\lambda_k$, contributes to 
    $\lambda_j$ integration. Here $D_k$ is the local mean level spacing at eigenvalue 
$\lambda_k$. As a consequence, an approximation of repulsion term 
$|\lambda_k-\lambda_j| \approx  D_k$ alongwith the relation 
 $\sum_{k=1}^N U_{nk} U^*_{mk}=\delta_{mn}$ (due to unitary nature of $U$) gives  
\begin{eqnarray}
\sum_{j=1;\not=k}^N {(U_{nj} U^*_{mj})^r \over (\lambda_k-\lambda_j)^{s}}=
{[\delta_{mn}-U_{nk} U^*_{mk}]^r \over (N-1)^{r-1} D_k^s}
\end{eqnarray}
Here $r=0,1$ only. 
$\chi=1$, $D_k$ is the local mean level spacing at eigenvalue $\lambda_k$.

(ii) for $\mu > \zeta_k^d$, the significant contribution comes from 
     the regions of $\lambda_j$ where 
$|\lambda_j-\lambda_k| \sim D_k [\zeta_k^d/\mu]^{1/2}$. Here again, 
as a typical $|U_{nj}|^2 \sim \zeta^{-d} < 1$, the entire region of $U_j$ can 
contribute to the integral. Consequently one can approximate 
\begin{eqnarray}
\sum_{j=1;\not=k}^N {(U_{nj} U^*_{mj})^r \over (\lambda_k-\lambda_j)^{s}}=
\left({\mu\over \zeta_k^d}\right)^{s/2}  
{[\delta_{mn}-U_{nk} U^*_{mk}]^r \over (N-1)^{r-1} D_k^s}
\end{eqnarray}
 (One may also consider the contribution from regions where 
$|U_{nj}|^2 < (\mu |U_{nk}|^2 D_k^2)^{-1}$ however it is weaker than the above).

By substituting approximations (A3,A4) in eq.(A1), $Q_{mn;k}^{rs}$ can be written as 
 (for $r=0,1$ only):

\begin{eqnarray}
 Q_{mn;k}^{rs} \approx  \chi^{s/2} {(\delta_{mn}-z_{mk}^* z_{nk})^r \over (N-1)^{r-1} D^s} 
P_{N1}(Z_k,e_k) 
\end{eqnarray}
where $\chi=1$ for $\mu< \zeta_k^d$ and $\chi=\mu/\zeta_k^d$ for $\chi>\mu/\zeta_k^d$.

	The integral $G_r$ (see eq.(30) can also be rewritten in terms of 
$\rho(H)$ and can similarly be approximated: 

\begin{eqnarray}
G_r(x,e) \equiv  \sum_{j;j\not=k}\int \delta^{\beta}_x 
\delta_e {|U_{nj}|^{2r} \over (\lambda_k-\lambda_j)^{2} } 
\rho \; {\rm d}H  
\end{eqnarray}
The dominant contribution in this case comes from those regions 
of integration over  $U_j$ and $\lambda_j$ 
which lead to ${\tilde R} \equiv \mu |U_{nk}|^2 \sum_{j} |\lambda_k-\lambda_j|^2 
|U_{nj}|^2$  present in the exponent of $\rho$. (Note, unlike the dominating term $R$ in 
$Q_{mn;k}^{rs}$ case, ${\tilde R}$ contains only a single component of the $k^{\rm th}$ 
eigenfunction, namely, $U_{nk}$, and, the latter takes a fixed value $x/{\sqrt N}$.) 
Consequently, for a given $Y$, $G_r$ depends on the mutual competition between 
$\mu$ and $x$. Reasoning as in the case of $Q_{mn;k}^{rs}$, $G_r$ can be approximated 
as

\begin{eqnarray}
G_r &\approx & \mu \chi_0 (N-1)^{1-r}(N-|x|^2)^r P_{11}(x)/D^2  
\qquad\qquad  
\end{eqnarray} 

with $\chi_0=\mu^{-1}$ for $\mu |x|^2 < 1$ and 
$\chi_0 \sim   |x|^2$ for $\mu |x|^2 > 1$.

\section{Effect of  Matrix Elements Perturbations on Eigenvalues and Eigenfunctions}
 
         Consider the perturbation of a Hermitian matrix $H$ with matrix elements 
$H_{kl} \equiv \sum_{s=1}^2 (i)^{s-1} H_{kl;s}$, eigenvalues $\lambda_n$ and 
eigenfunctions $U_n$, $n=1,2,..N$. By using the eigenvalue equation 
$\sum_{m} H_{nm} U_{mj}=\lambda_n U_{nj}$ along with the ortho-normal condition 
on eigenvectors i.e. $\sum_j U_{nj} U^*_{mj}=\delta_{mn}$, it can be shown that 

 \begin{eqnarray}
 {\partial \lambda_n \over\partial H_{kl;s}}
  &=& 2 g_{kl}^{-1} U_{kn} U_{ln}  \nonumber \\
 {\partial U_{nj} \over\partial H_{kl;s}} &=&
 {i^{s-1}\over g_{kl}} \sum_{m\not=j}
 {1\over {\lambda_j -\lambda_m}}
 U_{nm}(U^*_{km}U_{ln} +  (-1)^{s+1}U^*_{lm} U_{kj}). 
 \end{eqnarray}
The details of the steps used in derivation of eq.(B1) can be found in \cite{ps1}.  
	
 	The set of equations (B1) can further be used to show following relations: 

 \begin{eqnarray}
 \sum_{k,l,s ;k\le l}  
{\partial \lambda_n \over\partial H_{kl;s}} H_{kl;s} &=&  \lambda_n \\
 \sum_{k;l,s; k\le l} {\partial U_{nj} \over\partial H_{kl;s}} H_{kl;s} &=& 0, \\ 
 \end{eqnarray}

and, 
 \begin{eqnarray}
 \sum_{k,l,s; k\le l} {g_{kl}\over 2}  
{\partial^2 U_{nj} \over\partial H_{kl;s}^2} &=&
 - \sum_{m\not=j}{U_{nj} \over (\lambda_j -\lambda_m)^2}\\
 \sum_{k,l,s; k\le l} g_{kl} {\partial \lambda_i \over\partial H_{kl;s}}
 {\partial U_{nj} \over\partial H_{kl;s}} &=&  0 \\
 \sum_{k,l,s; k\le l} {g_{kl}} {\partial U_{ni} \over\partial H_{kl;s}} 
 {\partial U_{pj} \over\partial H_{kl;s}} &=&
 - \beta {U_{ni } U_{nj}\over (\lambda_i -\lambda_j)^2} (1-\delta_{ij})\delta_{np} \\
 \sum_{k,l,s; k\le l} {g_{kl}} {\partial U_{ni} \over\partial H_{kl;s}} 
 {\partial U^*_{pj} \over\partial H_{kl;s}} &=&
  \beta \sum_{m \not=j} {U_{nm } U^*_{pm}\over (\lambda_j -\lambda_m)^2} \delta_{ij} 
 \end{eqnarray}

 \end{appendix}

 \section{Figure Caption}
 \noindent Fig. 1. 
 Distribution $P_u(u')$  with $u'=[{\rm ln}u-\langle {\rm u} \rangle]/ 
{\langle {\rm ln}^2 u \rangle}$  of the local intensity of an eigenfunction 
near band center for  AE$_{t}$ (Cubic lattice of linear size $L=13$, with 
hard wall boundary conditions, random hopping and time-reversal symmetry) 
and its BE and PE analog. The parts (a) and (b) of the figure show short 
and long range behavior of the distribution, respectively. The analogs are 
obtained by the relation 
$I_{2,a}^{typical}/F_a(0)=I_{2,b}^{typical}/F_b(0)=I_{2,p}^{typical}/F_p(0)$.
For $N=2197$, we find $I_{2,a}^{typical}= 0.018, I_{2,b=0.02}^{typical}= , 
I_{2,p=0.4}^{typical}=0.02$, 
and, $F_a(0)=0.26, F_{b=0.02}(0)=(\pi)^{-1/2}, F_{p=0.4}(0)=0.39$.


\noindent Fig. 2 The local intensity distribution for  AE$_{nt}$ 
(Cubic lattice of linear size $L=13$, with periodic boundary conditions,
non-random hopping and no time-reversal symmetry) and its BE and PE analogs.
In this case,  $I_{2,a}^{typical}= 0.013, I_{2,b=0.03}^{typical}= ,
I_{2,p=0.4}^{typical}=0.00045$, and, $F_a(0)=0.016, F_{b=0.03}(0)=(\pi)^{-1/2},
F_{p=0.4}(0)=$ for $N=2197$. Other details are same as in figure 1. 


\noindent Fig. 3.
 Distribution $P(I'_2)$ of the rescaled inverse participation ratio 
 $I'_2= {\rm ln}[I_2/I_2^{typ}]$
 for  AE$_{t}$ (same as in Figure 1(a)) and its BE and PE analog:
(a) short range behavior (lin-lin plot), (b) tail behavior (lin-log plot). 
 Here the BE and PE analogs  are obtained by the relation 
$\Lambda_{I,a}={\Lambda}_{I,b}={\Lambda}_{I,p}$. This gives a 
BE analog of AE different from the figure 1 although PE analog remains 
unaffected; the reason lies in almost similar mean level density 
behavior near band-center for the AE, PE cases.

\noindent Fig. 4.
 Distribution $P(I'_2)$ of the rescaled inverse participation ratio
 $I'_2$ for  
 AE$_{nt}$ (same as in Figure 1(b)) and its BE and PE analog. 
The other details are same as in figure 3.
Note the BE analog of AE, PE in this case is different from the 
figure 2.

\noindent Fig. 5. 
 Distribution $P(w')$ of the spatial correlation $w'={\rm ln} w=
{\rm ln} |z_{1n} z_{Nn}|^2$ between two  points belonging to opposite 
end of the sample: 
(a) short range behavior (lin-lin plot), (b) tail behavior (lin-log plot). 
The cases compared here are AE$_{t}$ 
(same as in Figure 1(a)) and its BE and PE analog (obtained by the 
relation $\Lambda_{w,a}={\Lambda}_{w,b}={\Lambda}_{w,p}$). 
Again the BE analog of AE, PE in this case turns out to be different 
from the figure 1 but same as in the figure 3.   


\noindent Fig. 6.
 Distribution $P(w')$ of the spatial correlation $w'$ 
for AE$_{nt}$ (same as in figure 2) and its BE and PE analog; 
other details are same as in figure 5.
Here the BE and PE analogs  are obtained by the relation
$\Lambda_{w,a}={\Lambda}_{w,b}={\Lambda}_{w,p}$. The BE analog of
AE, PE in this case is different from the figure 2 but same as in 
the figure 4.

\noindent Fig. 7.
 Distribution $P(S)$ of the nearest-neighbor spacing distribution $S$ of 
the eigenvalues, with (a), (b) showing short and long range behavior, 
respectively, for  AE$_{t}$ (same as in Figure 1) and its BE and PE analog.
Here the BE and PE analogs  are obtained by the relation
$\Lambda_{e,a}={\Lambda}_{e,b}={\Lambda}_{e,p}$. Note the BE analog of
AE, PE in this case is different from the figure 1 but same as in the 
figures 3 and 5.
                                                                                                                             
\noindent Fig. 8.
 Distribution $P(S)$ of the nearest-neighbor spacing distribution $S$
for the case AE$_{nt}$ (same as in Figure 2) and its BE and PE analog 
(other details same as in figure 7). 

 

\begin{figure}
\begin{minipage}[b]{0.5\linewidth}
\centering
\includegraphics[scale=0.45]{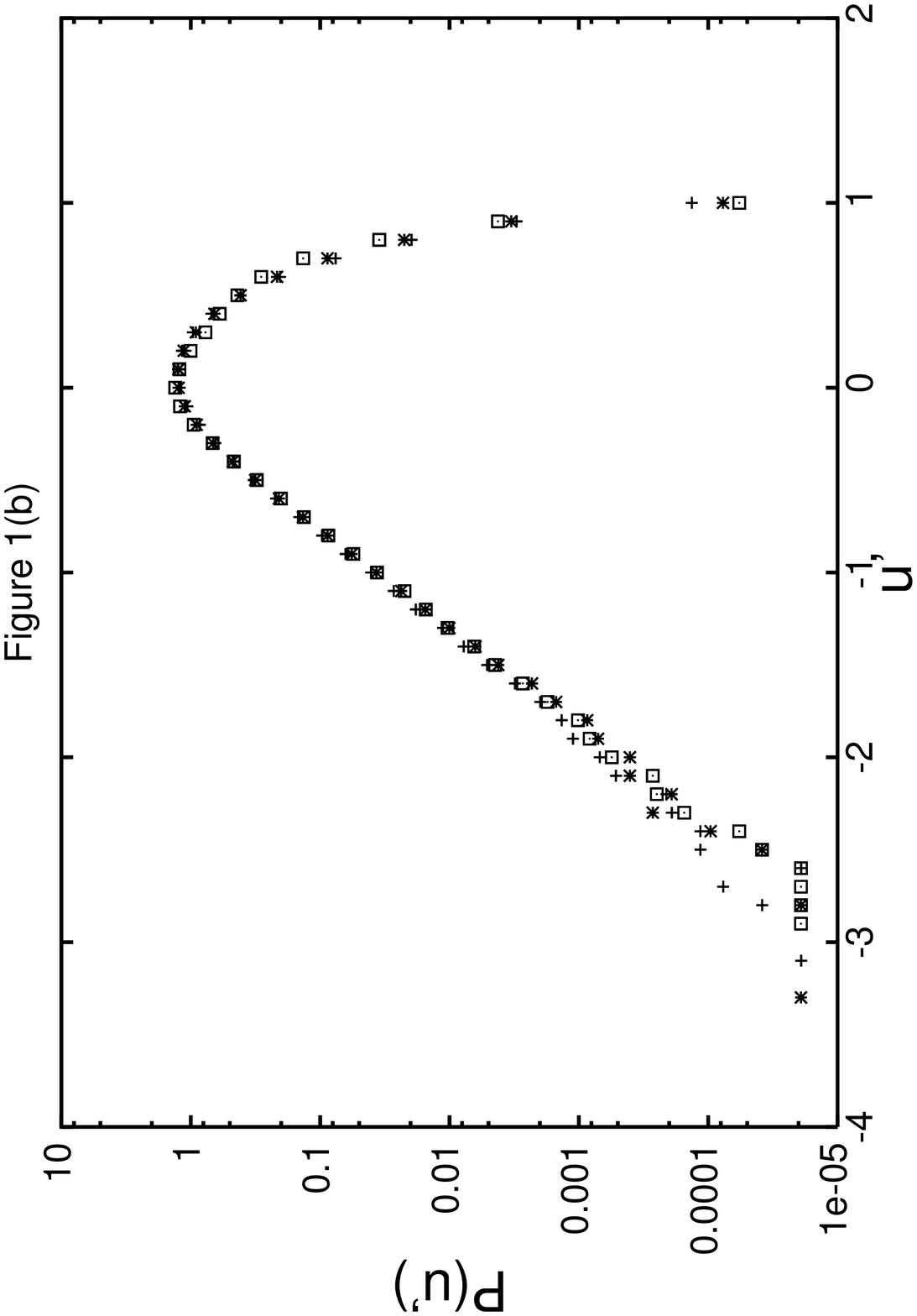}
\end{minipage}
\begin{minipage}[b]{0.5\linewidth}
\centering
\includegraphics[scale=0.45]{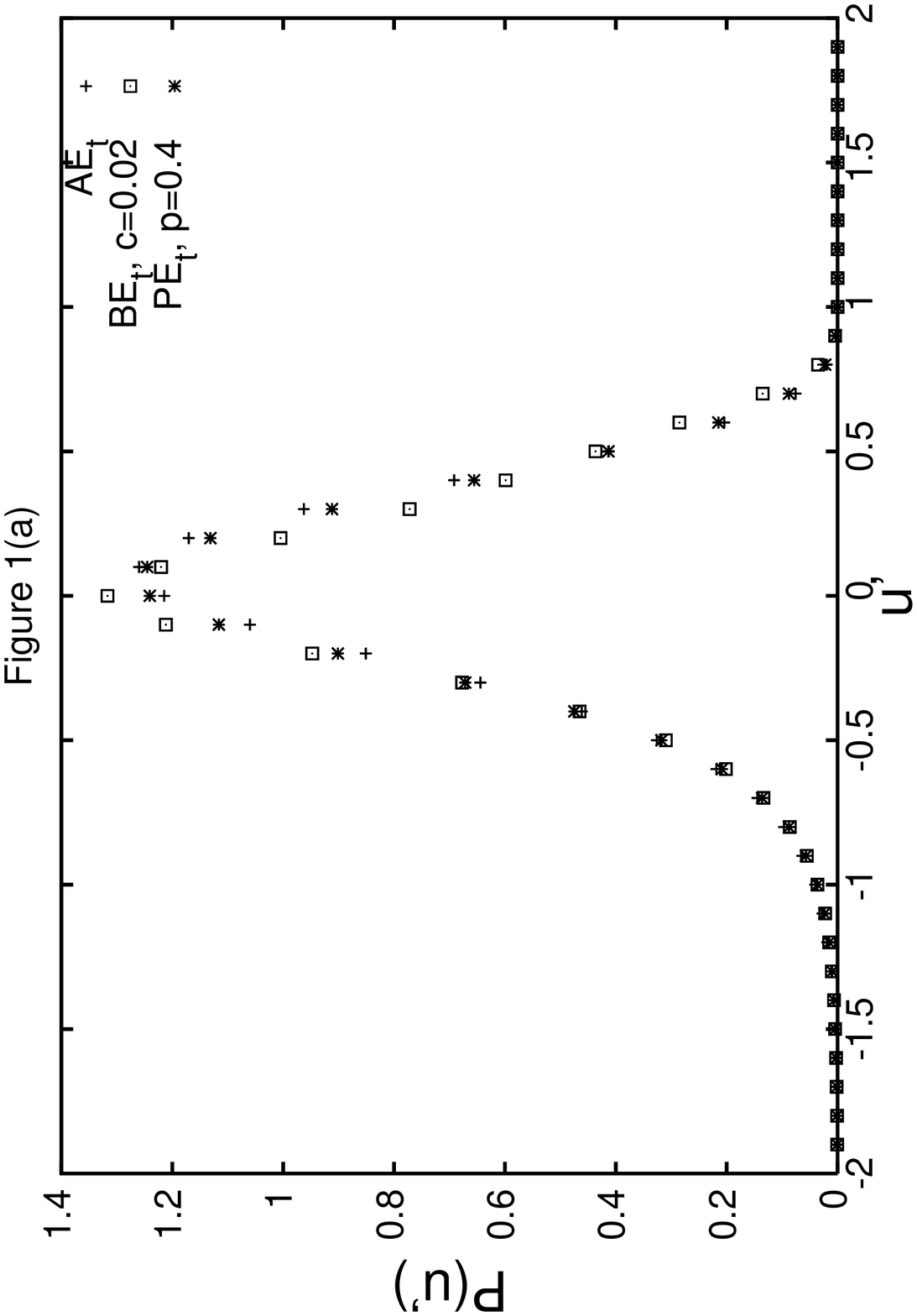}
\end{minipage}
\end{figure}

\begin{figure}
\begin{minipage}[b]{0.5\linewidth}
\centering
\includegraphics[scale=0.45]{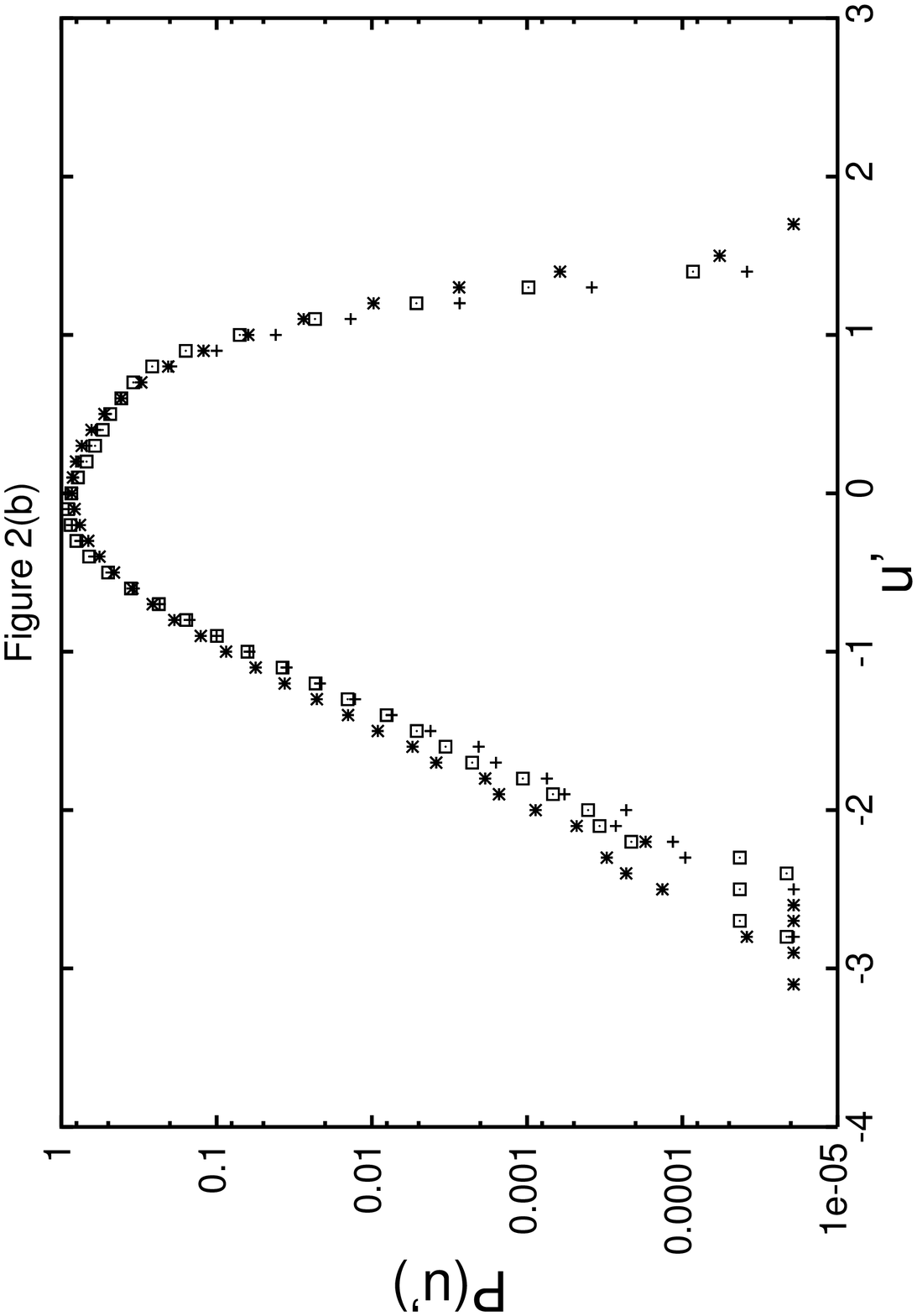}
\end{minipage}
\begin{minipage}[b]{0.5\linewidth}
\centering
\includegraphics[scale=0.45]{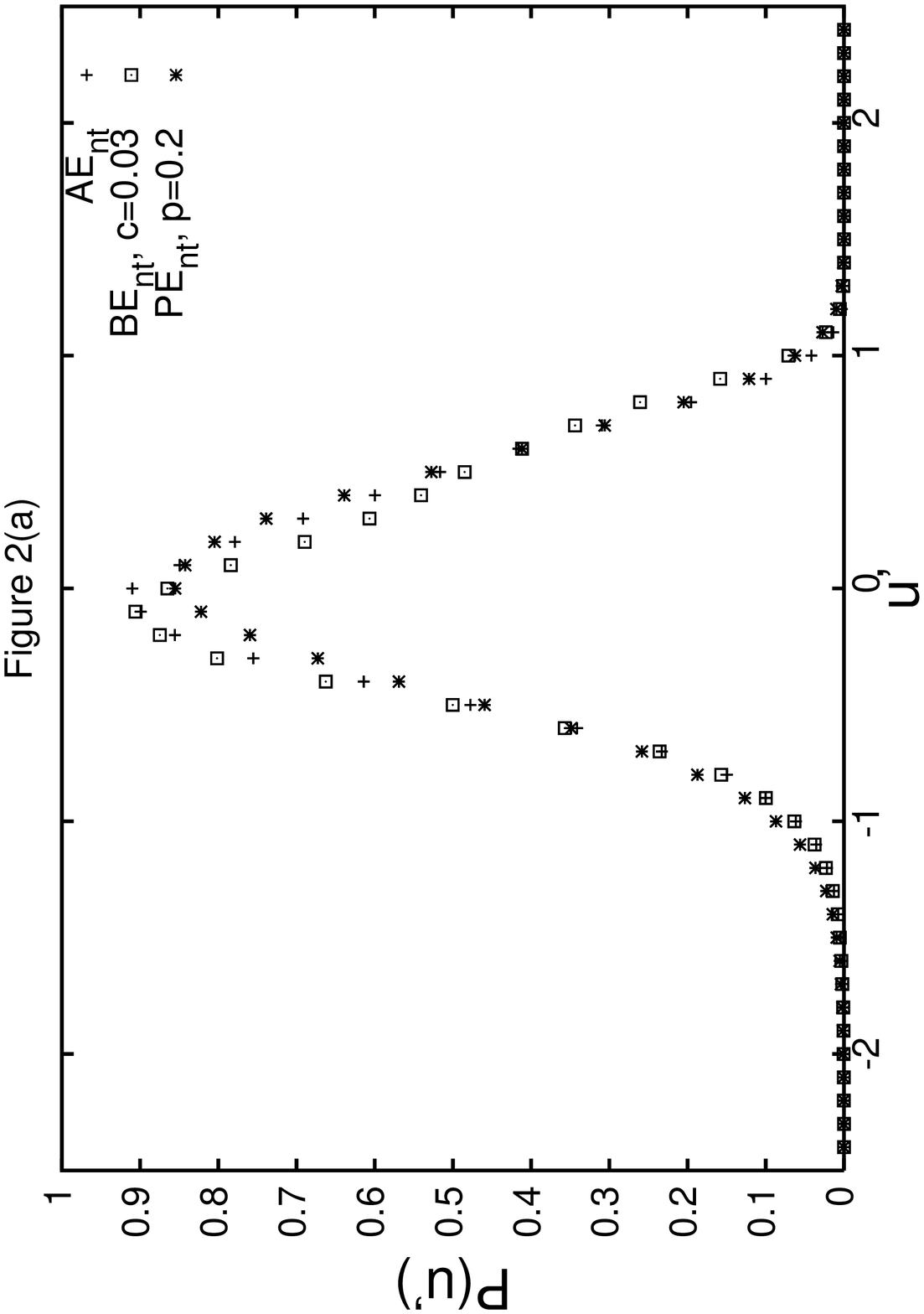}
\end{minipage}
\end{figure}

\begin{figure}
\begin{minipage}[b]{0.5\linewidth}
\centering
\includegraphics[scale=0.45]{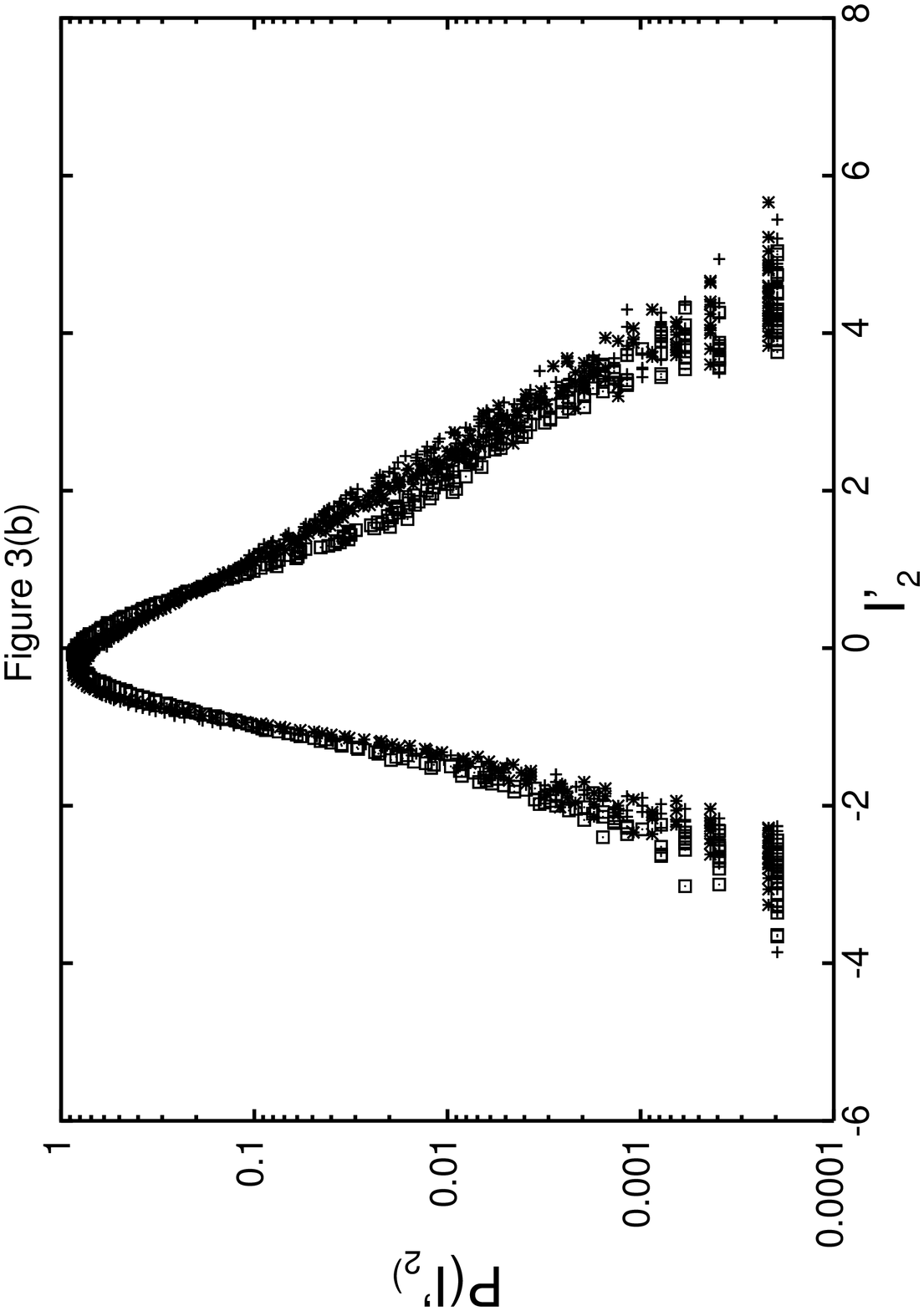}
\end{minipage}
\begin{minipage}[b]{0.5\linewidth}
\centering
\includegraphics[scale=0.45]{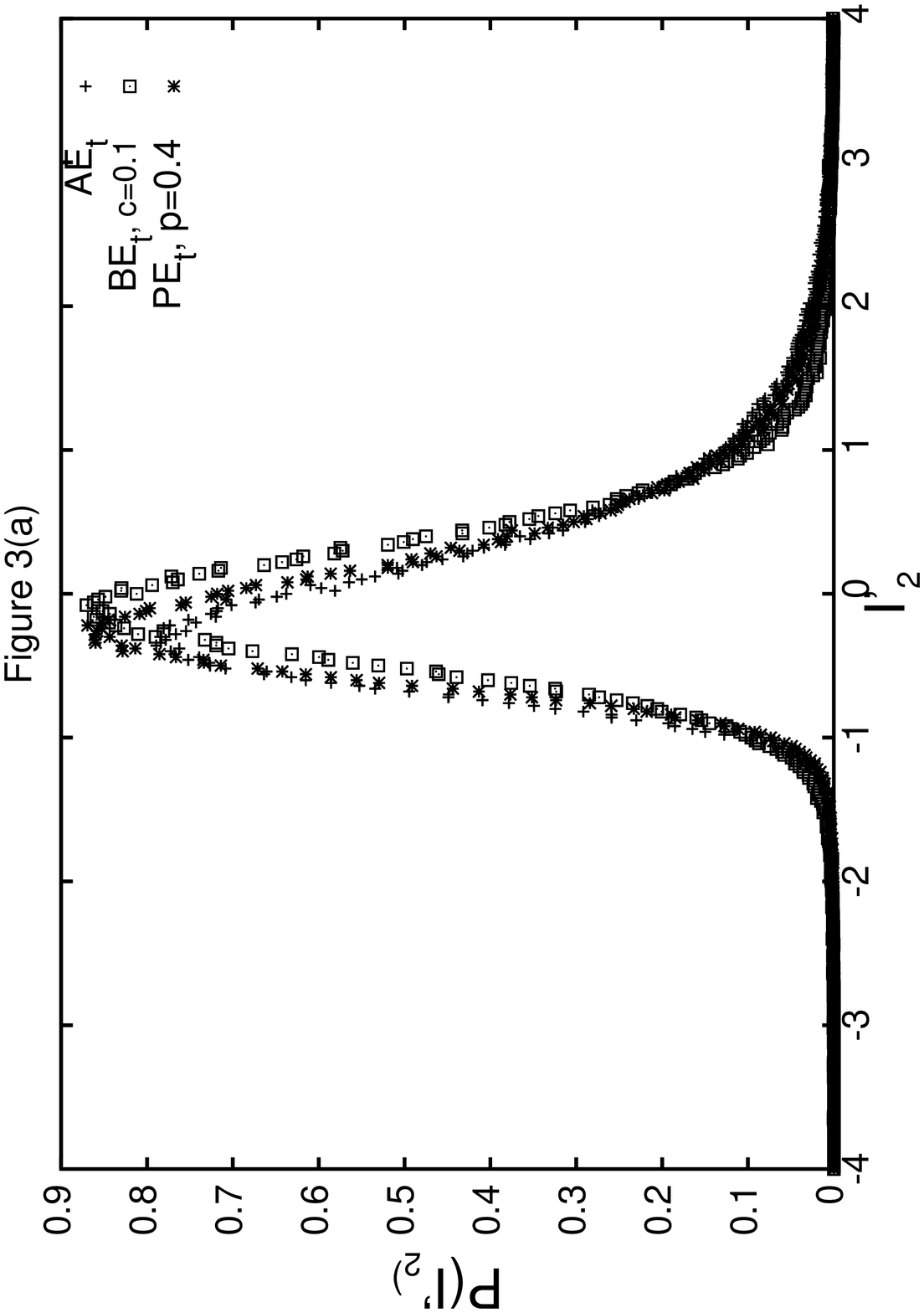}
\end{minipage}
\end{figure}

\begin{figure}
\begin{minipage}[b]{0.5\linewidth}
\centering
\includegraphics[scale=0.45]{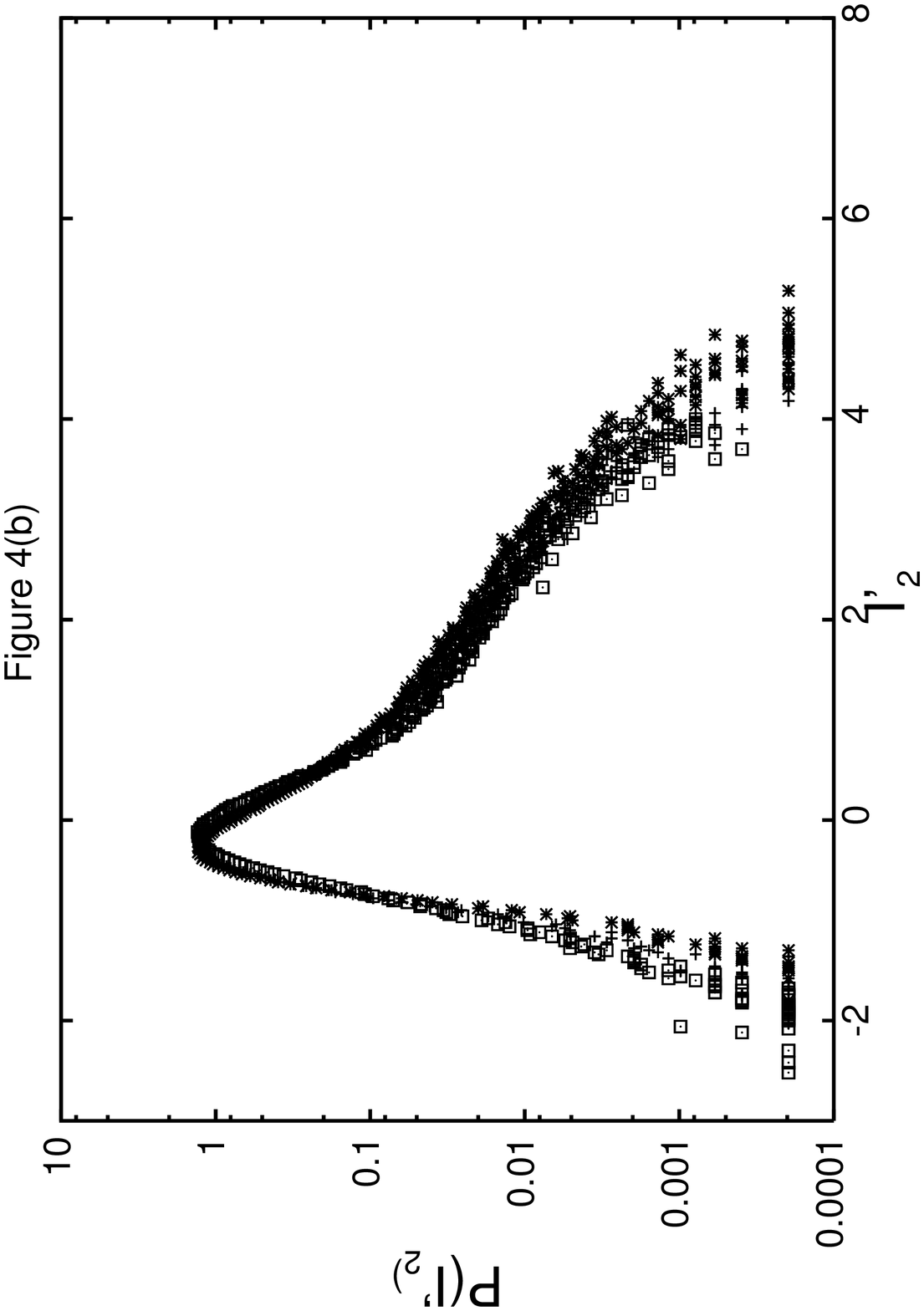}
\end{minipage}
\begin{minipage}[b]{0.5\linewidth}
\centering
\includegraphics[scale=0.45]{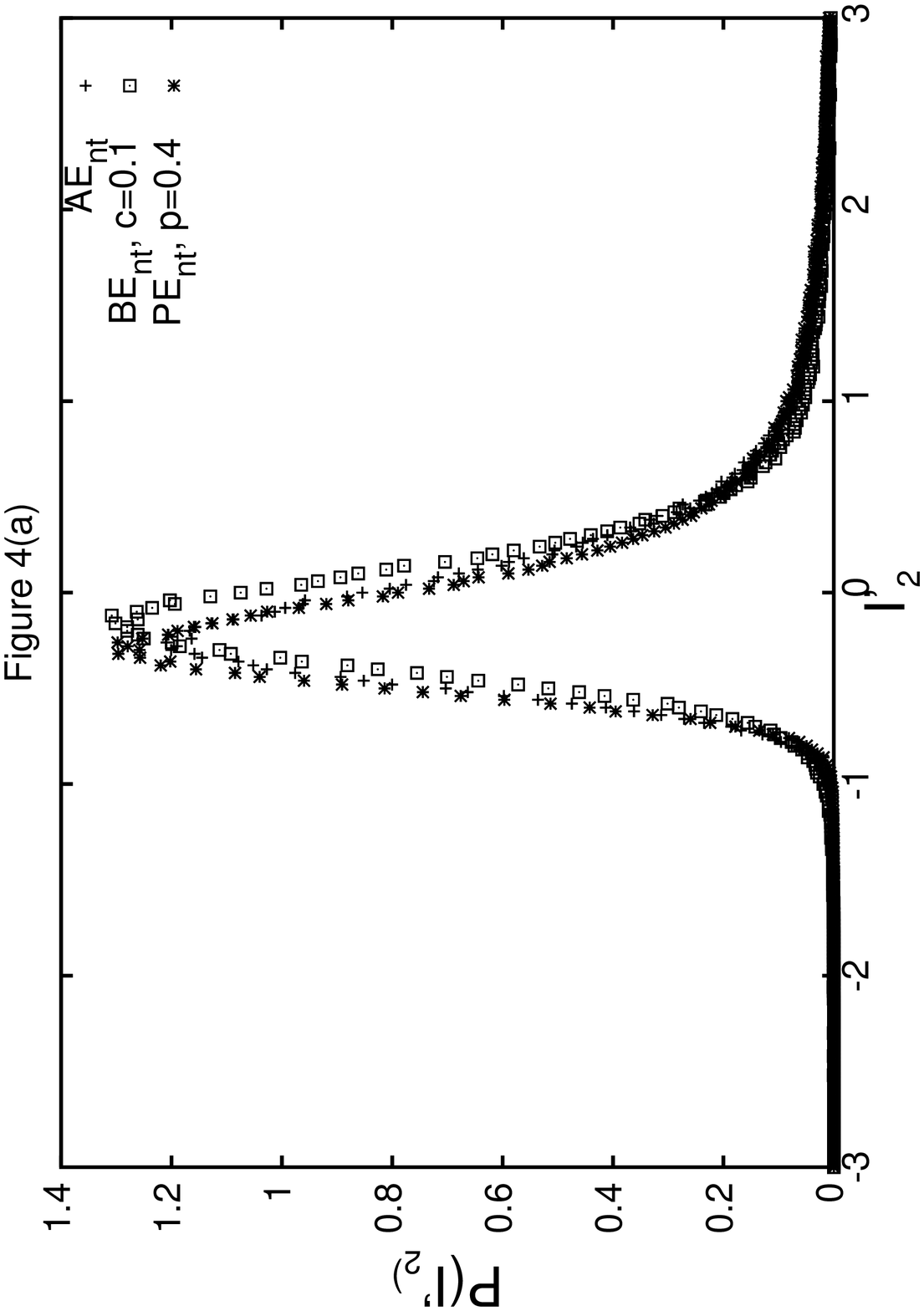}
\end{minipage}
\end{figure}

\begin{figure}
\begin{minipage}[b]{0.5\linewidth}
\centering
\includegraphics[scale=0.45]{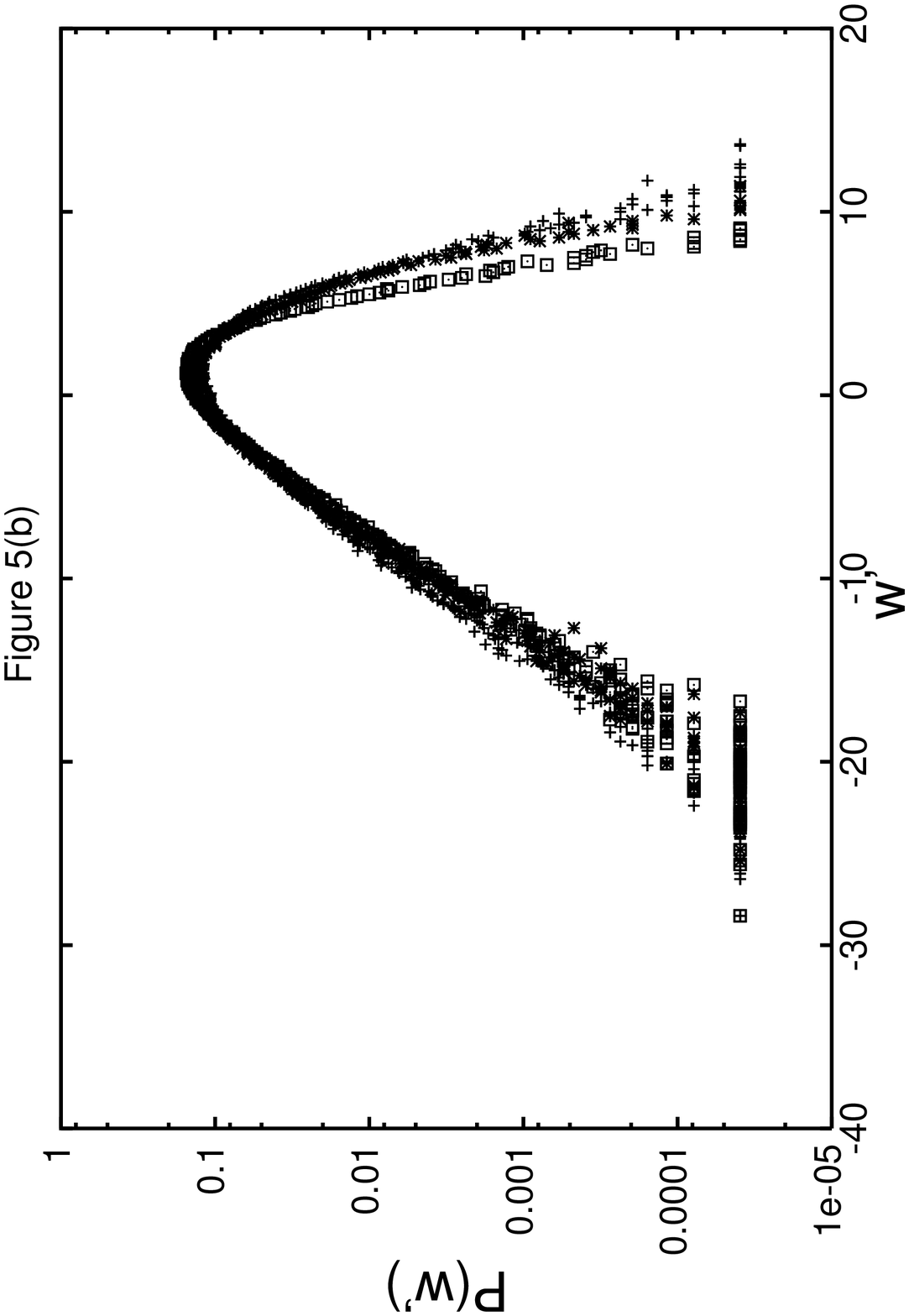}
\end{minipage}
\begin{minipage}[b]{0.5\linewidth}
\centering
\includegraphics[scale=0.45]{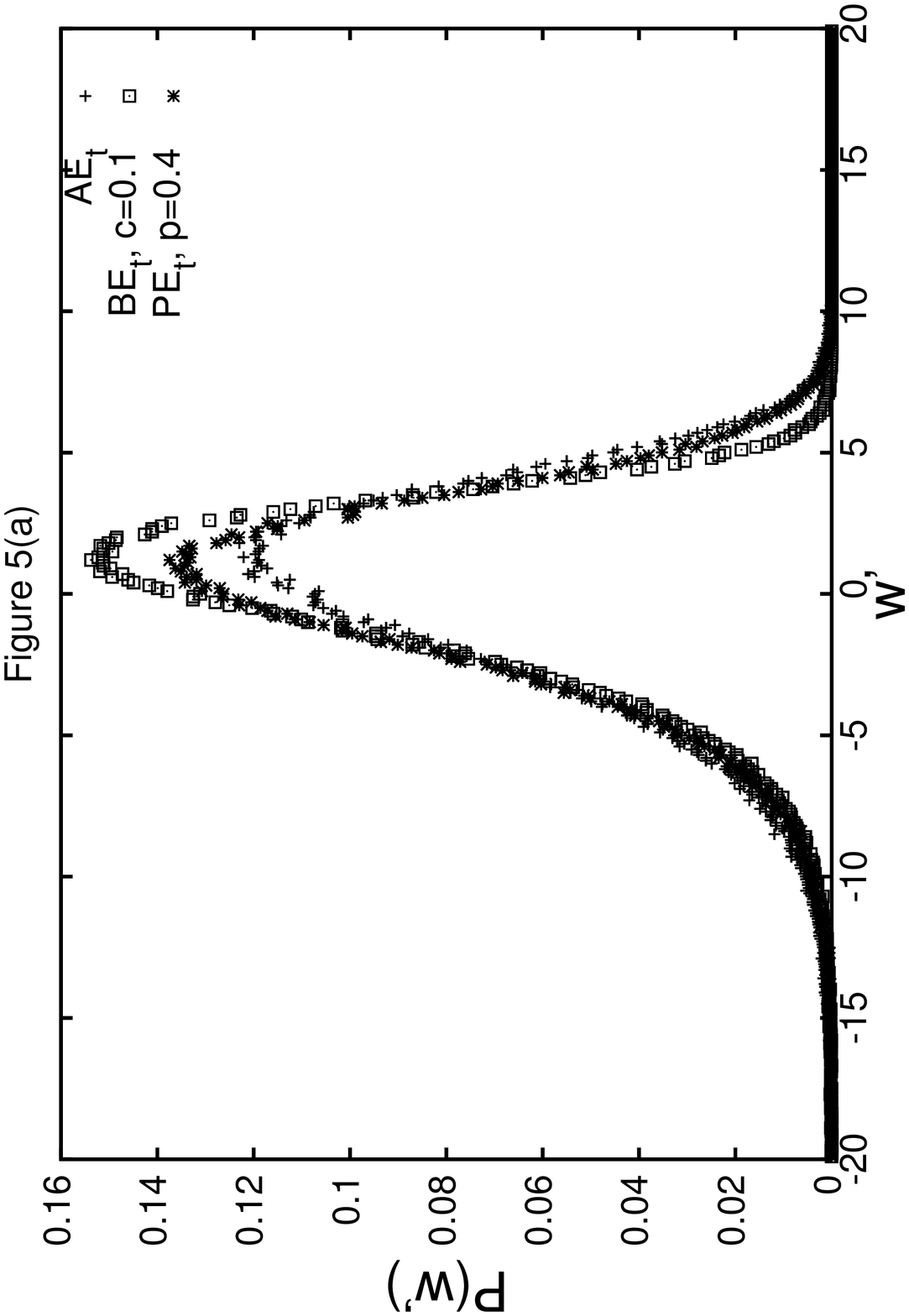}
\end{minipage}
\end{figure}

\begin{figure}
\begin{minipage}[b]{0.5\linewidth}
\centering
\includegraphics[scale=0.45]{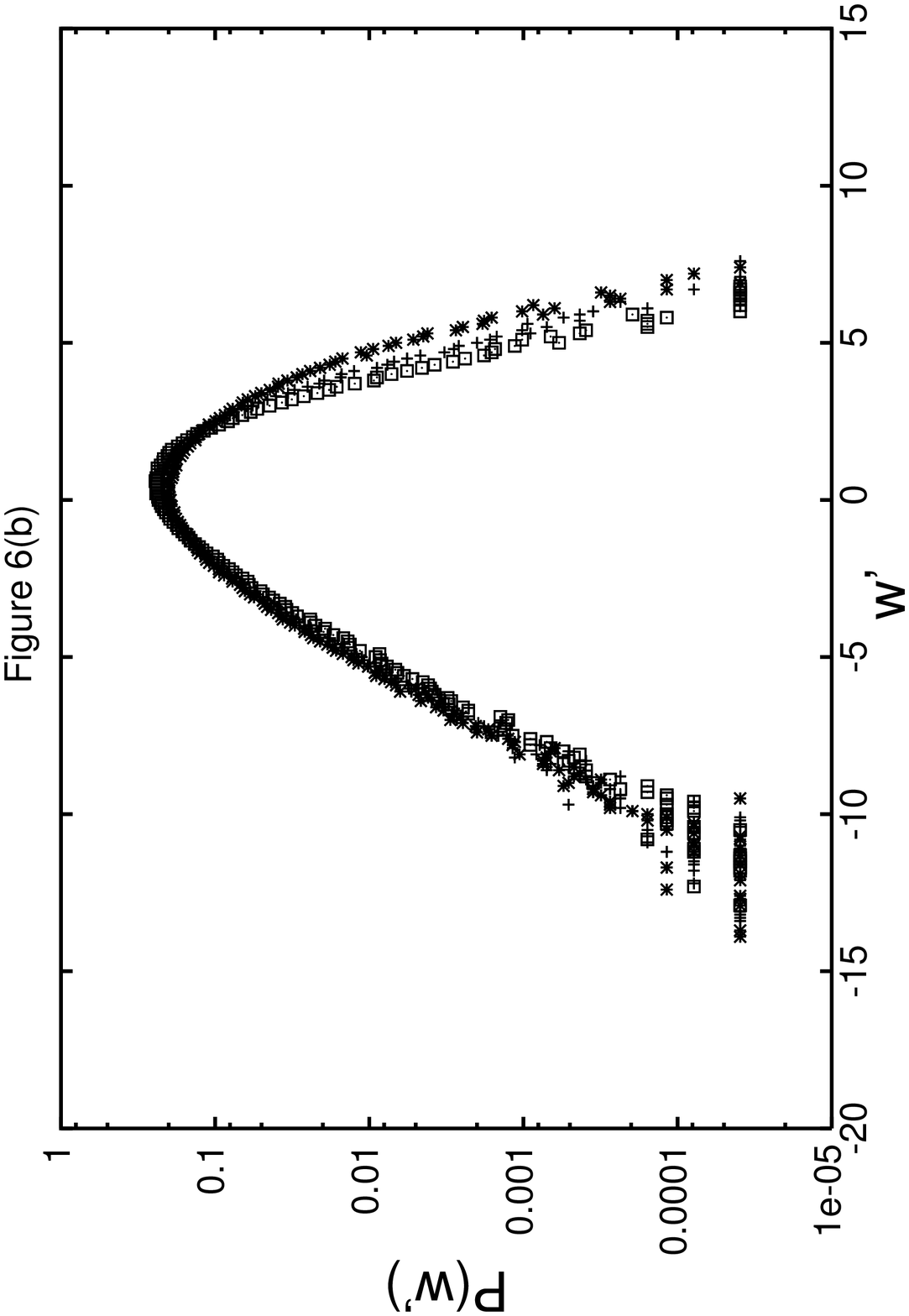}
\end{minipage}
\begin{minipage}[b]{0.5\linewidth}
\centering
\includegraphics[scale=0.45]{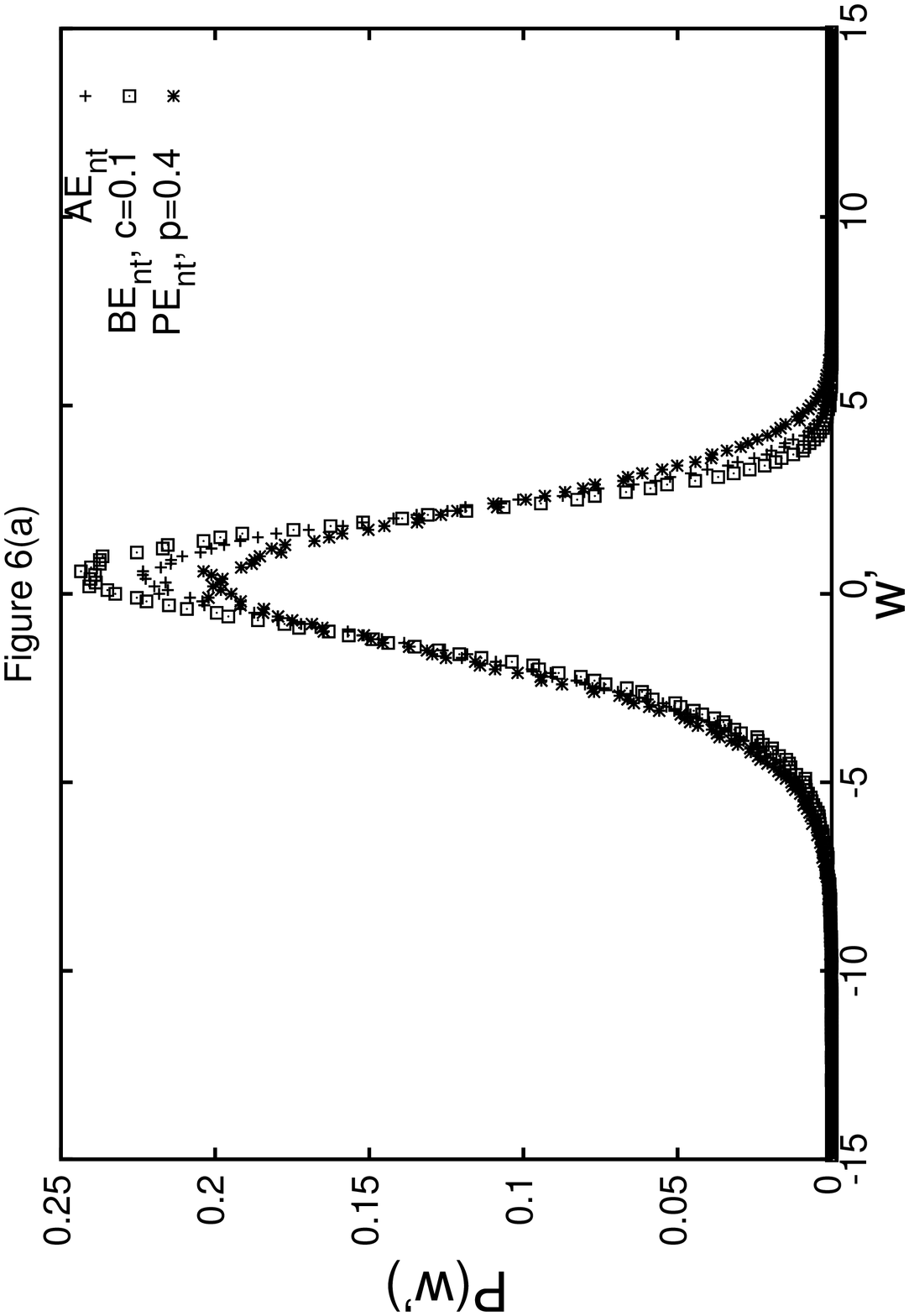}
\end{minipage}
\end{figure}

\begin{figure}
\begin{minipage}[b]{0.5\linewidth}
\centering
\includegraphics[scale=0.45]{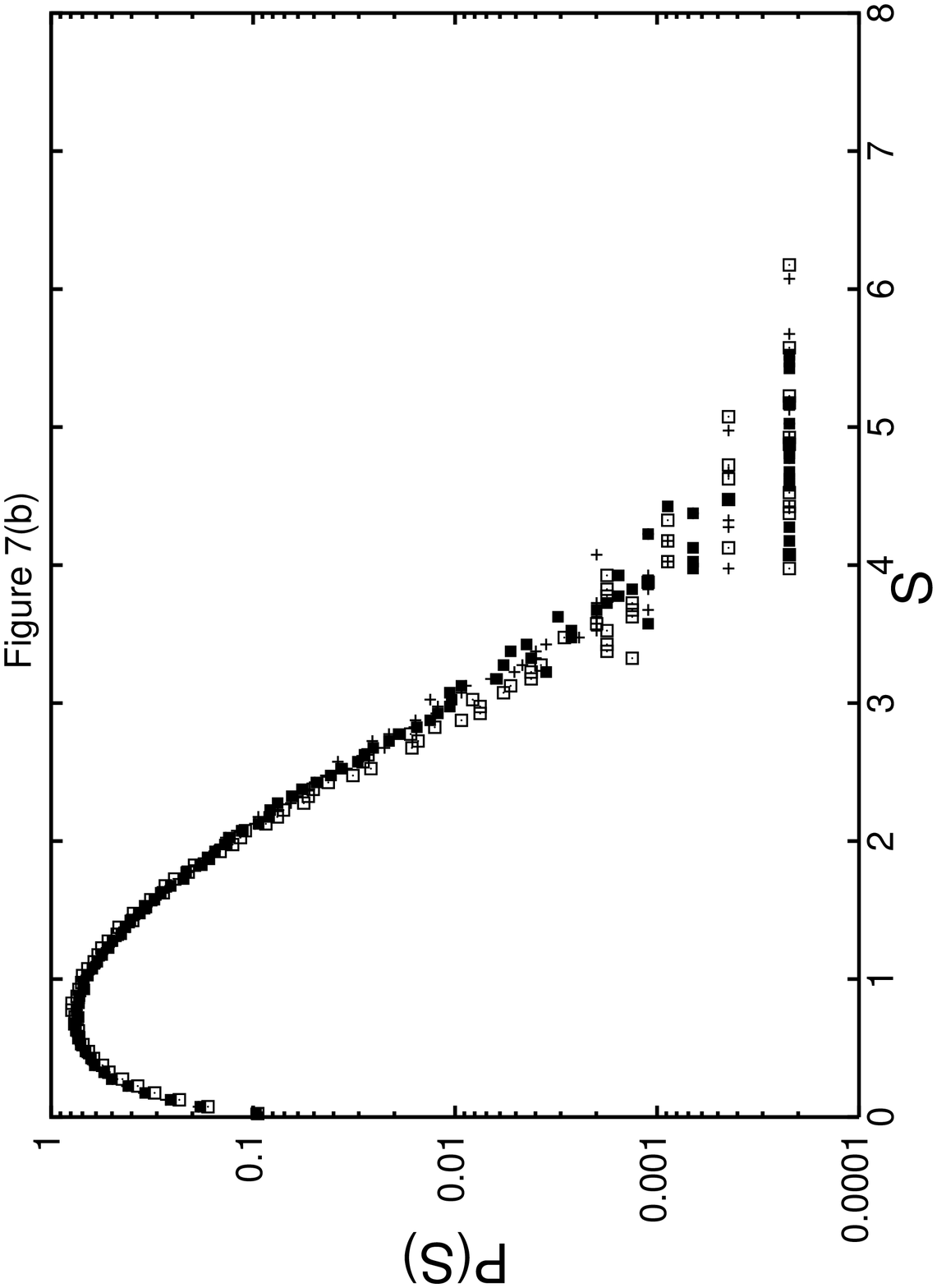}
\end{minipage}
\begin{minipage}[b]{0.5\linewidth}
\centering
\includegraphics[scale=0.45]{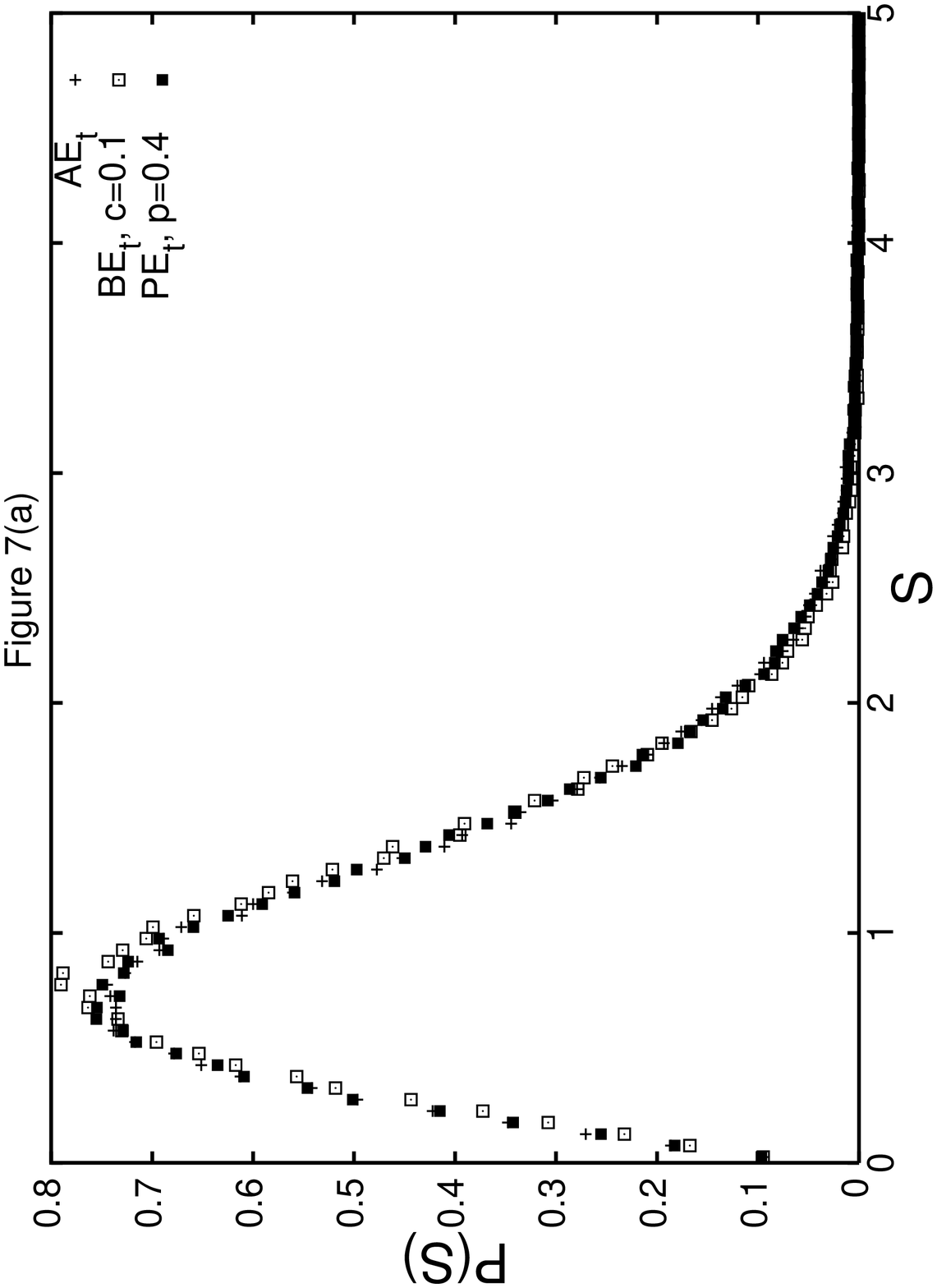}
\end{minipage}
\end{figure}

\begin{figure}
\begin{minipage}[b]{0.5\linewidth}
\centering
\includegraphics[scale=0.45]{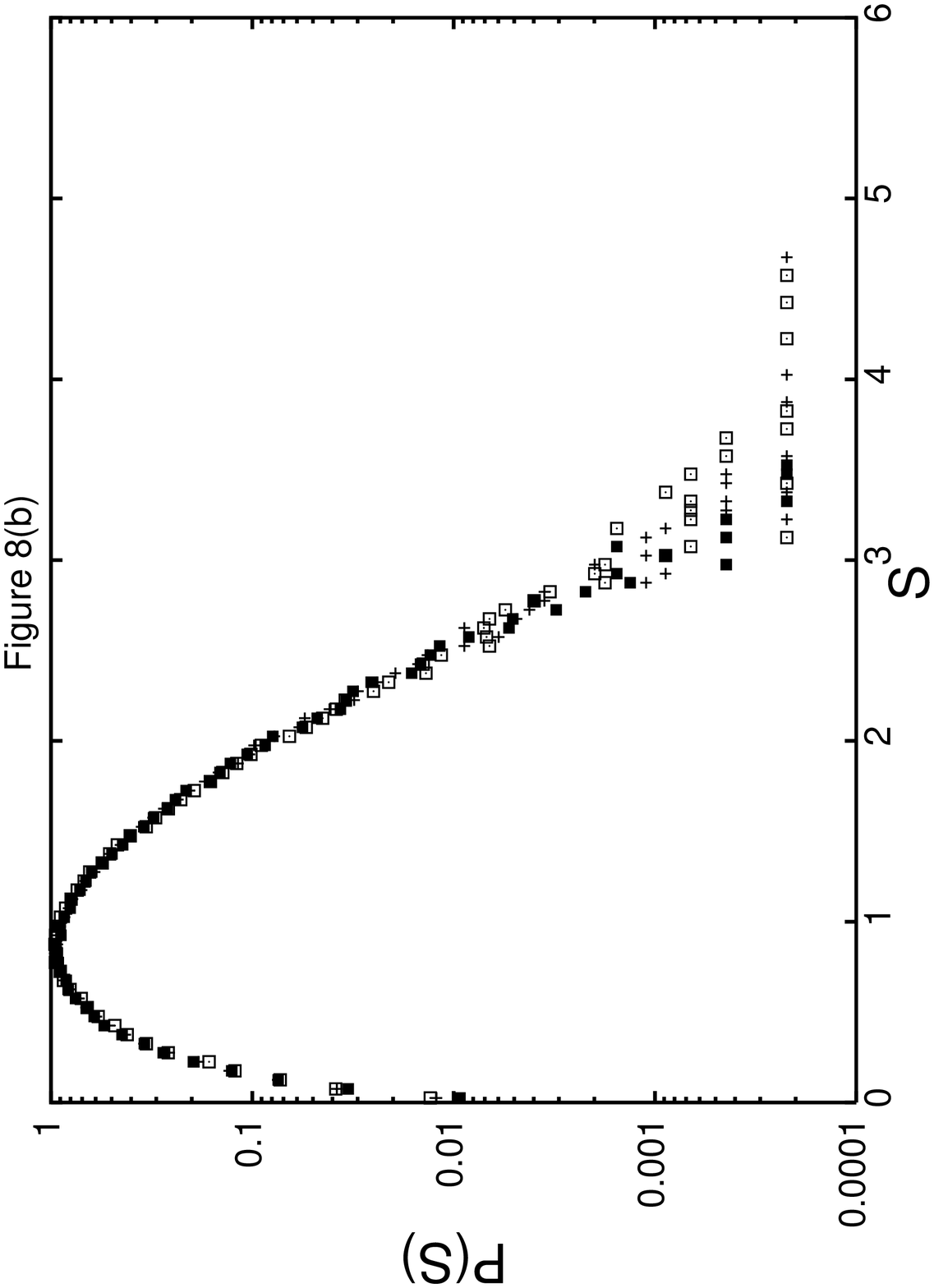}
\end{minipage}
\begin{minipage}[b]{0.5\linewidth}
\centering
\includegraphics[scale=0.45]{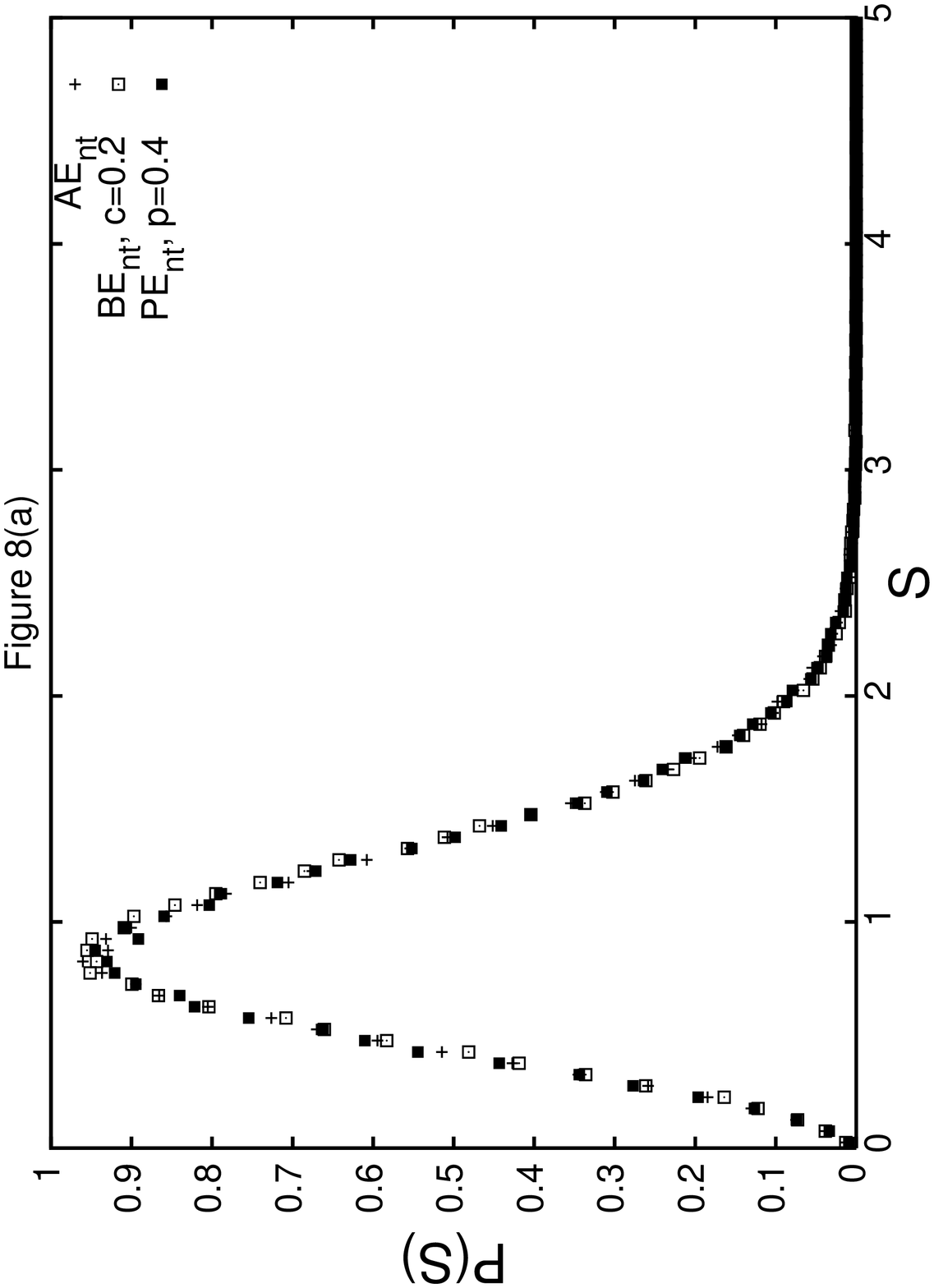}
\end{minipage}
\end{figure}

 \end{document}